\documentclass[%
 reprint,
 amsmath,amssymb,
 aps,
prb,
noeprint,
]{revtex4-1}

\usepackage[version=3]{mhchem} 
\usepackage{graphicx}
\usepackage{dcolumn}
\usepackage{bm}
\usepackage{hyperref}
\hypersetup{colorlinks=true, citecolor=blue, urlcolor=blue, linkcolor=blue}

\usepackage{color}

\usepackage{booktabs}
\usepackage[caption=false]{subfig}

\begin{document}

\preprint{APS/123-QED}

\title
{Influence of crystal structure on charge carrier effective masses in \ce{BiFeO3}}

\author{J. Kane Shenton}
\email{john.shenton@mat.ethz.ch}
\altaffiliation{
Institute of High Performance Computing, 1 Fusionopolis Way, \#16-16 Connexis North, Singapore 138632}
\altaffiliation{Now at Materials Theory, ETH Zurich,
Wolfgang-Pauli-Strasse 27,
CH-8093 Z\"urich, Switzerland}
\affiliation
{Department of Physics \& Astronomy,\\ 
University College London,\\ 
Gower St, London, WC1E 6BT}

\author{David R. Bowler}
\altaffiliation{International Centre for Materials Nanoarchitectonics (WPI-MANA), National Institute for Materials Science (NIMS), 1-1 Namiki, Tsukuba, Ibaraki 305-0044, Japan}
\altaffiliation{
Department of Physics \& Astronomy, University College London, Gower St, London, WC1E 6BT}
\affiliation{London Centre for Nanotechnology, \\
17-19 Gordon St, London, WC1H 0AH}
\author{Wei Li Cheah}
\affiliation{Institute of High Performance Computing,\\
1 Fusionopolis Way, \#16-16 Connexis North, Singapore 138632}



\date{\today}

\begin{abstract}
    Ferroelectric-based photovoltaics have shown great promise as a source of renewable energy, thanks to their in-built charge separation capability, yet their efficiency is often limited by low charge carrier mobilities.
    In this work, we compare the photovoltaic prospects of various phases of the multiferroic material \ce{BiFeO3} by evaluating their charge carrier effective masses from first-principles simulations. We identify a tetragonal phase with the promising combination of a large spontaneous polarisation and relatively light charge carriers.
    From a systematic investigation of the octahedral distortions present in \ce{BiFeO3}, we clarify the relationship between structure and effective masses. This relationship is explained in terms of changes to the orbital character and overlap at the band edges that result from changes in the geometry.
    Our findings suggest some design principles for how to tune effective masses in \ce{BiFeO3} and similar materials through the manipulation of their crystal structures in experimentally accessible ways.
\end{abstract}

\maketitle


\section{Introduction}

The field of ferroelectric (FE) photovoltaics, dating back to 1956 \cite{Chynoweth1956}, experienced something of a renaissance in 2009 with the discovery of the switchable diode and bulk photovoltaic effects in the multiferroic material, bismuth ferrite (\ce{BiFeO3}; BFO) \cite{Choi2009}.
The bulk photovoltaic effect, observed in BFO, arises from the absence of inversion symmetry in the crystal structure of the room temperature, $R\,3\,c$ phase \cite{Fridkin2001}. This asymmetry results in a giant spontaneous polarisation ($\sim100\mu C/cm^2$)\cite{Lebeugle2007,Shvartsman2007} that aids in charge separation in a photovoltaic device.


Despite the intrinsic charge separation ability, FE materials such as BFO generally suffer from low charge carrier mobilities, leading to high recombination losses and limited device efficiency \cite{Yuan2014,Butler2015,Huang2010}. Increasing the mobilities in such FE materials would therefore lead to enhanced photovoltaic device efficiency. 

Although mobility is a macroscopic quantity, it is the mobility associated with the electronic band edges that is of particular relevance to recombination rates.
Since mobility is directly related to the dispersion of these bands, a material's charge carrier mobilities can be altered by modifying the curvature of the bands near the Fermi level. Manipulation of the crystal structure, for example by strain engineering, is one such route to tuning the curvature of electronic bands. Strain engineering, together with chemical doping, have been widely exploited in the semiconductor industry to control mobility in silicon-based devices \cite{Sun2007}.
In BFO, a wide range of crystal structures can be stabilised through strain \cite{Sando2014,Sando2016} and interface \cite{Yu2012,Kim2013b} engineering. 
Of these structures, those that lack a centre of inversion symmetry are of significant interest for photovoltaic applications. There has been a great deal of interest in the tetragonal (spacegroup: $P\,4\,m\,m$) and tetragonal-like phases of BFO thin films, in particular, due to their giant spontaneous polarisation of \emph{ca.} 150 $\mu C/cm^2$.\cite{Sando2014,Sando2016}

How do the mobilities of the charge carriers compare across the BFO phases?
In this work, we present an investigation within density functional theory (DFT) of the electronic properties of several FE and non-FE phases of BFO.
We begin by considering the experimentally accessible ground-state $R\,3\,c$ phase and the higher temperature orthorhombic $P\,n\,m\,a$ and cubic $P\,m\,\bar{3}\,m$ phases. We also consider the prototypical tetragonal $P\,4\,m\,m$ phase, accessible via epitaxial strain. \cite{Sando2014,Sando2016} We further consider the theoretical $R\,\bar{3}\,c$ phase, which is similar to the $R\,3\,c$ phase but without the polar distortion that gives the latter its spontaneous polarisation.
We compare the charge carrier effective masses ($m^{*}$), which are inversely proportional to the mobilities, across these phases.
We will show that the carriers in the room temperature FE ($R\,3\,c$) phase have considerably larger $m^{*}$ compared to some non-FE phases. Nonetheless we find that the tetragonal phase of BFO, which has a large spontaneous polarisation, has relatively low electron and hole effective masses.

In order to explain the differences in $m^{*}$ across the BFO phases, we systematically study the geometric transformations that map the BFO phase with the lowest $m^{*}$, to the phase with the highest $m^{*}$. The effects of these transformations on the $m^{*}$ are explained in terms of changes to the orbital character and overlap at the band edges.
Previous works \cite{Sando2016} have indicated that, with a judicious choice of substrate material, a BFO lattice with the desired spontaneous polarisation and $m^{*}$ can be fabricated. Our results therefore provide insight into the rational design of materials with optimum properties, particularly for applications in light harvesting.



\section{\label{sec:calc_details}Computational details\protect}

We consider the $R\,3\,c$, $P\,n\,m\,a$, $P\,4\,m\,m$, $P\,m\,\bar{3}\,m$ and $R\,\bar{3}\,c$ phases of BFO.
Simulations of these phases were performed using DFT as implemented in the Vienna \emph{ab initio} Simulation Package (VASP Version 5.4.1) \cite{Kresse1993,Kresse1994,Kresse1996a,Kresse1996b}.
We employ the version of the generalised gradient approximation parameterised by Perdew, Burke and Ernzerhof (PBE) \cite{Perdew1996} as the exchange-correlation functional, with an effective Hubbard correction, $\mathrm{U_{eff}}$, applied to the Fe $d$ orbitals using the method of Dudarev \emph{et al.} \cite{Dudarev1998}. This PBE+U method has been employed in this work to account for the known failures of standard local density and generalised gradient approximations to accurately describe the strong correlations in transition metal oxides \cite{Dudarev1998}.
We use a $\mathrm{U_{eff}}$ of 4 eV for all calculations as this value has been found to best capture the electronic structure of BFO, particularly the orbital characters at the valence band maximum ($\mathrm{VB_{max}}$) and conduction band minimum ($\mathrm{CB_{min}}$) \cite{Shenton2017}.

All of the calculations were carried out using the projector-augmented plane-wave method \cite{Blochl1994b,Kresse1999b} and a plane-wave cut-off energy of 520 eV, treating explicitly 15 electrons for Bi ($5d^{10}6s^{2}6p^{3}$), 14 for Fe($3p^{6}3d^{6}4s^{2}$), and 6 for O ($2s^{2}2p^{4}$).\footnote{The Bi, Fe and O PAWs are dated: 6$\mathrm{^{th}}$ Sept. 2000, 2$\mathrm{^{nd}}$ Aug. 2007 and 8$\mathrm{^{th}}$ Apr. 2002 respectively}
The 40-atom pseudocubic (pc) unit cell setting was consistently adopted for all but the effective mass calculations in section \ref{subsec:geom_trans}. The pc unit cell setting allows us to capture the C- or G-type antiferromagnetism (AFM) across all of the BFO phases and allows for a more straightforward comparison of the structures than the various primitive cells. For reasons of efficiency, the effective mass calculations in section \ref{subsec:geom_trans} were performed using a 10-atom rhombohedral cell setting.
Brillouin zone integrations for the relaxations and static calculations were performed on $\Gamma$-centred Monkhorst-Pack (MP) \cite{Monkhorst1976} grids: $9\times9\times9$ for the pc unit cells and $11\times11\times11$ for the rhombohedral unit cells. 
Density of states (DOS) calculations, requiring finer sampling of the Brillouin zone, were performed using a $\Gamma$-centred $11\times11\times11$ MP grid in the pc unit cell setting.
We relaxed the low-energy phases of BFO such that all force components were less than 5 meV/\AA.
The unit cell shape and sizes of these phases were optimised such that all stress components were smaller than 0.06 GPa.
All of the structure files and raw data are available in the Supplementary Information (SI) at Ref.\citenum{Shenton2018}

We use the modern (Berry's phase) theory of polarisation (MTP) \cite{King-Smith1993,Resta1994} to calculate the spontaneous polarisation of each structure.
As noted by Neaton \emph{et al.} \cite{Neaton2005}, much care is needed when performing these calculations, especially in systems such as BFO for which the spontaneous polarisation, $P_s$, is of the same order of magnitude as the quantum of polarisation, $Q$. We follow the procedure outlined in Ref. \citenum{Neaton2005}.

We calculate the hole and electron effective masses for each of the considered BFO phases as follows.
The curvature\footnote{For a thorough discussion of alternative effective mass definitions, we refer the reader to the recent work of Whalley \emph{et al.} \cite{Whalley2019}} effective mass of a given band, $n$, at a particular location in reciprocal space, $\boldsymbol{k}$, is a $\mathrm{3\times 3}$ tensor quantity whose magnitude in a given direction is inversely proportional to the band curvature in that direction. $m^{*}$ can therefore be defined as:
\begin{equation}
    \left(\frac{1}{m^*}\right)_{ij} = %
    \frac{1}{\hbar^2}\frac{\partial^2 E_n(\boldsymbol{k})}{\partial k_ik_j}, %
    i,j=x,y,z,
\end{equation}
where $E_n(\boldsymbol{k})$ is the energy dispersion relation for the $n^{th}$ band, and $i$ and $j$ represent reciprocal space components.
To obtain $m^{*}$, we first compute the full band-structure along a path of high symmetry. We then identify the location of the $\mathrm{VB_{max}}$ and the $\mathrm{CB_{min}}$ in reciprocal space. The band curvatures at these points correspond to the hole effective mass, $m^{*}_{h}$, and the electron effective mass, $m^{*}_{e}$, respectively.
Having identified these $k$-points of interest, we employ the method and code outlined in Ref. \citenum{Fonari2012} to obtain the full $m^{*}$ tensors.
In brief, the method involves generating a fine mesh around the $k$-point of interest, calculating the energy eigenvalues, and using a finite difference method to build up the tensor of second derivatives.
The dependence of $m^{*}$ on the spacing of this mesh was investigated, and spacings of less than 0.05 bohr$^{-1}$ were found to give consistent results.
We calculate the eigenvalues of the effective mass tensor, which correspond to the effective masses along the principle directions.
For the purposes of this work, in which we consider the photovoltaic prospects of different phases, we envisage a device oriented in such a way as to take advantage of any anisotropy in the $m^{*}$ tensor. With that in mind, we compare the smallest effective masses of each structure, with the full $m^{*}$ tensors presented in the SI.

To make the calculations tractable, we limit ourselves to a collinear treatment of spin, thus neglecting the long-wavelength ($\sim 620$ \AA) spiral spin structure found experimentally \cite{Sosnowska1982b}. 
All of the low-energy phases in the present work, with the exception of the $P\,4\,m\,m$ phase, are found to adopt a G-type AFM ordering, consistent with previous work \cite{Dieguez2011}. The C-type AFM ordering was found to be slightly lower (6 meV /f.u.) in energy than the G-type ordering for the $P\,4\,m\,m$ phase, again in agreement with previous work \cite{Dieguez2011,Hatt2010}. 
A change from G-type to C-type ordering in the $P\,4\,m\,m$ phase had little effect on the character or curvature of the band edges.

\section{\label{sec:results}Results\protect}
\subsection{\label{subsec:relaxed_struct}Effective masses of the phases of BFO\protect}

    Table \ref{tab:relaxed} shows the calculated $m^{*}$, polarisation and relative stability of the considered BFO phases.
    The results for the relative stability agree well with the literature, differing by at most 5\% with respect to the PBE+U ($\mathrm{U_{eff}}$=4 eV)  work of Di{\'{e}}guez \emph{et al.} \cite{Dieguez2011}.
    For the FE $R\,3\,c$ phase, the calculated $P_s$ agrees well with both experiment and theory, which report values of between 90-100 $\mu C /cm^2$ \cite{Neaton2005,Ederer2005b,Ravindran2006,Lebeugle2007}. 
    For the tetragonal $P\,4\,m\,m$ phase, we find a $P_s$ of 185 $\mu C /cm^2$, larger than those reported in previous theoretical (151-152 $\mu C /cm^2$ \cite{Ederer2005a, Dieguez2011} and experimental ($\sim130$ $\mu C /cm^2$ \cite{Zhang2011}) works.

    Comparing the charge carrier effective masses, reported in Table \ref{tab:relaxed}, we find the room-temperature FE phase of BFO ($R\,3\,c$) to have largest $m^{*}_{e}$, and the second largest $m^{*}_{h}$. 
    The tetragonal $P\,4\,m\,m$ phase, which is also ferroelectric, has an $m^{*}_{e}$ an order of magnitude smaller and an $m^{*}_{h}$ 20\% smaller than those of $R\,3\,c$. 
    The paraelectric $R\,\bar{3}\,c$ phase has an $m^{*}_{e}$ comparable to that of $P\,4\,m\,m$ and an $m^{*}_{h}$ comparable to that of $R\,3\,c$. Both $m^{*}$ are smallest in the paraelectric $P\,m\,\bar{3}\,m$ phase.
    From these results we therefore observe that the ferroelectric phases do not necessarily have lower $m^{*}$ than the paraelectric phases.

\begin{table}[]
    \centering
    \caption{\textmd{Computed effective masses in units of electron rest mass, $m_0$, spontaneous polarisation values and relative energies of five PBE+U relaxed phases of BFO. Note that energies, $E$, are taken relative to the most stable phase: $R\,3\,c$.}}
    \label{tab:relaxed}
    \begin{tabular}{l@{\hskip 0.5cm}r@{\hskip 0.5cm}r@{\hskip 0.5cm}r@{\hskip 0.5cm}r}
    \toprule[1pt]
     Spacegroup & $|m^{*}_{h}|$  & $m^{*}_{e}$ & Polarisation & $E - E_{R\,3\,c}$) \\
      & $(m_{0})$ & $(m_{0})$ & ($\mu C/cm^{2}$) & (meV per f.u.) \\
    \midrule[0.5pt]
     $P\,m\,\bar{3}\,m$ & 0.34 & 0.24 &  0.0   & 971 \\
     $R\,\bar{3}\,c$       & 0.63 & 0.37 &  0.0   & 268 \\
     $P\,4\,m\,m$         & 0.54 & 0.33 &185.3  &  84 \\
     $P\,n\,m\,a$          & 0.95 & 0.99 &  0.0    &  57 \\
     $R\,3\,c$               & 0.67 & 3.06 & 90.0   &   0 \\
    \bottomrule[1pt]
    \end{tabular}
\end{table}

\begin{figure*}
        \centering
        \subfloat[$R\,3\,c$]{\includegraphics[width=0.35\linewidth]{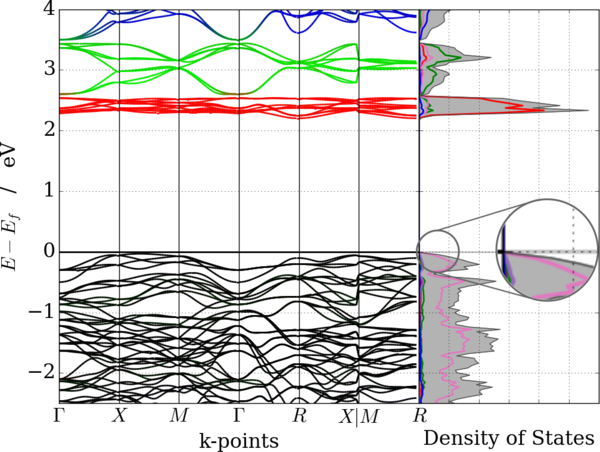}}\hspace{1em}%
        \subfloat[$P\,n\,m\,a$]{\includegraphics[width=0.35\linewidth]{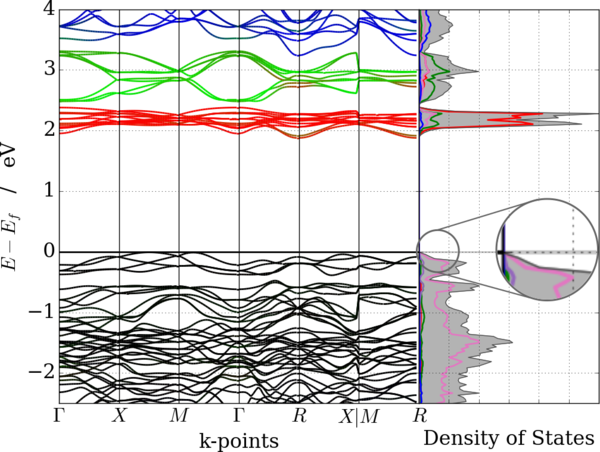}}

        \subfloat[$R\,\bar{3}\,c$]{\includegraphics[width=0.35\linewidth]{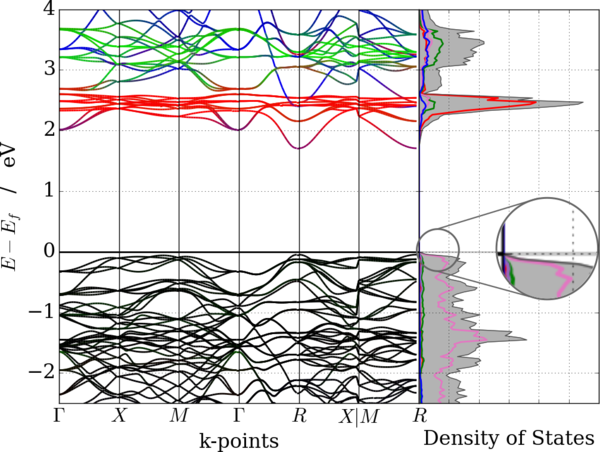}}\hspace{1em}%
        \subfloat[$P\, 4\, m\, m$ ]{\includegraphics[width=0.35\linewidth]{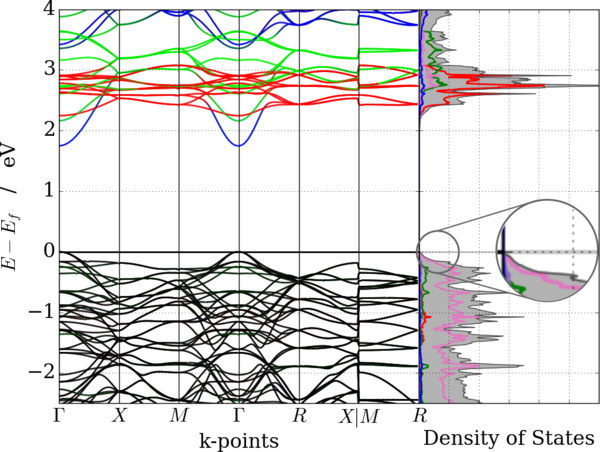}}

        \subfloat[$P\,m\,\bar{3}\,m$]{\includegraphics[width=0.35\linewidth]{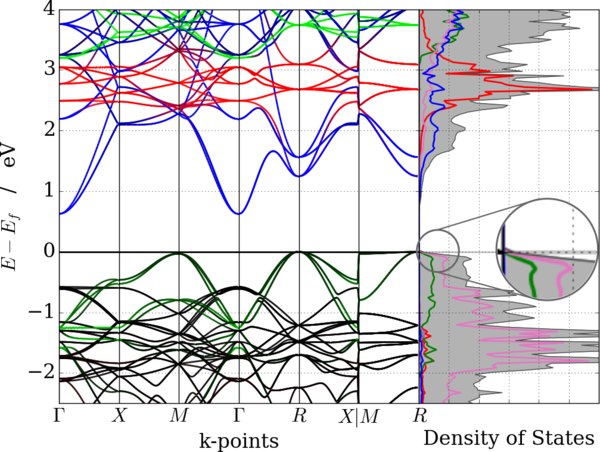}}\hspace{1em}%
        \subfloat{\includegraphics[width=0.35\linewidth]{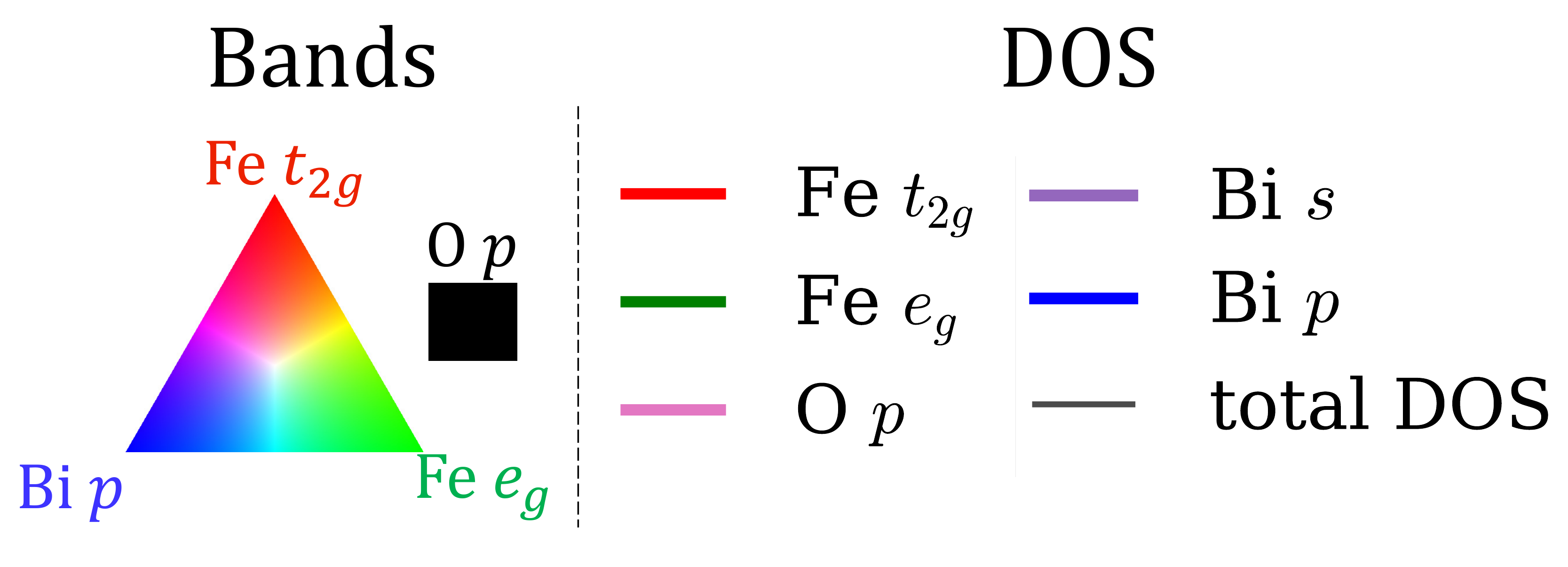}}
        \caption{Projected bands and DOS for the (a) $R\,3\,c$, (b) $P\,n\,m\,a$, (c) $R\,\bar{3}\,c$, (d) $P\,4\,m\,m$ and (e) $P\,m\,\bar{3}\,m$ phases of BFO, in their (pseudo-) cubic settings.
        The bands are coloured, at each $k$-point, based on wavefunction projections onto chosen orbitals. As indicated in the legend, red, green, blue and black represent projections onto Fe $t_{2g}$, Fe $e_{g}$, Bi $p$ and O $p$ states respectively. 
        }\label{fig:relaxed_pbands}
    \end{figure*}

To gain insight into the variation of $m^{*}$ with the phases, we examine and compare the electronic structures of the BFO phases, particularly at the band edges - the locations at which the $m^{*}$ are evaluated.
In Fig. \ref{fig:relaxed_pbands} the band structure and density of states (DOS), projected onto spherical harmonics, are shown for each phase of BFO studied here.
Considering the lowest lying conduction bands in order of decreasing $m^{*}_{e}$, i.e. from $R\,3\,c$, $P\,n\,m\,a$, $R\,\bar{3}\,c$, $P\,4\,m\,m$ to $P\,m\,\bar{3}\,m$, we see that the major contribution to the states at $\mathrm{CB_{min}}$ gradually changes from Fe $t_{2g}$ to Bi $p$.
In Fig. S1 of the SI \cite{Shenton2018}, we plot the corresponding Kohn-Sham (KS) orbitals. These orbital plots corroborate the spherical harmonic projections, indicating that Fe $t_{2g}$ states are the primary contributors to the $\mathrm{CB_{min}}$ for the $R\,3\,c$, $P\,n\,m\,a$ and $R\,\bar{3}\,c$ phases, while the Bi $p$ states make up the $\mathrm{CB_{min}}$ for the $P\,4\,m\,m$ and $P\,m\,\bar{3}\,m$ phases. Our results suggest that, at the $\mathrm{CB_{min}}$, the presence of Bi $p$ states leads to a lower $m^{*}_{e}$ than that of Fe $t_{2g}$ states.

\begin{figure*}
    \includegraphics[width=0.75\linewidth]{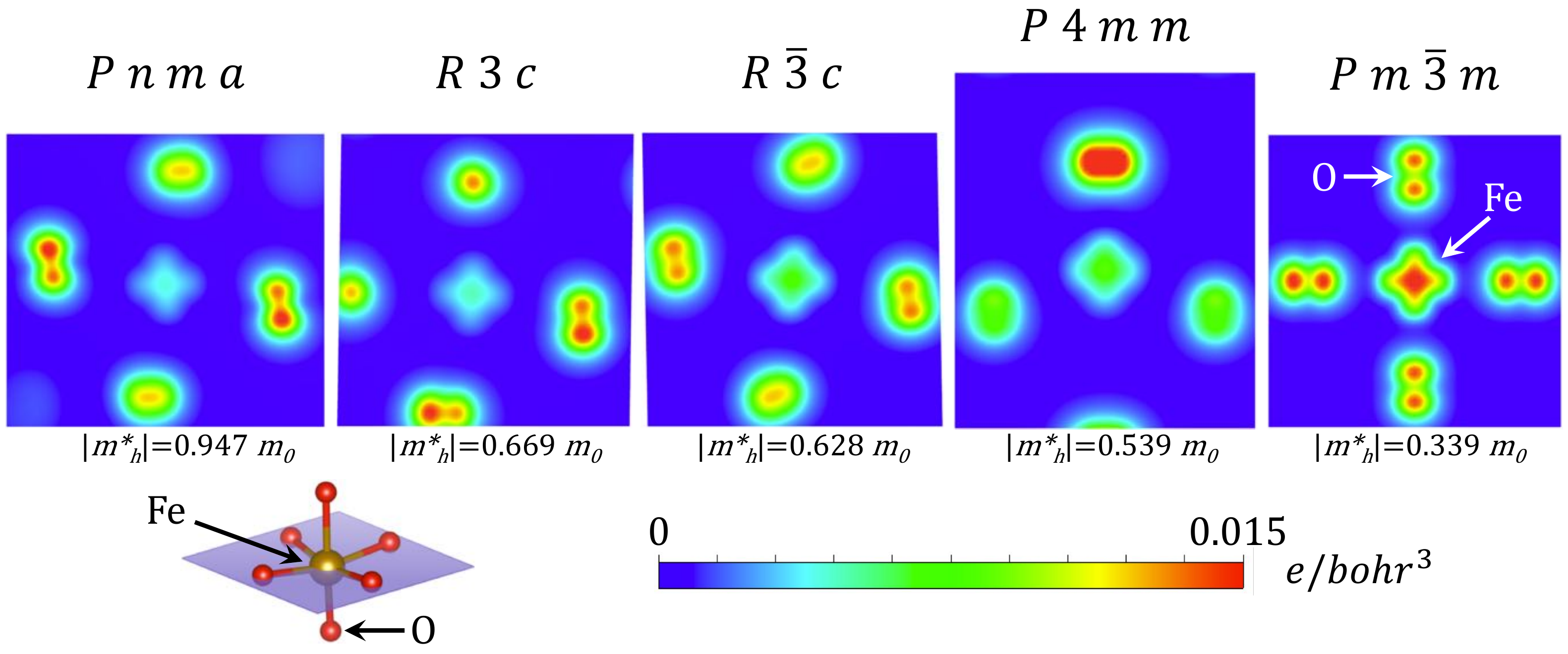}
    \caption{Kohn-Sham orbitals corresponding to top VB, summed over all $k$-points, for the five phases of BFO investigated. These are representative cross-sections through the FeO$_4$ plane in the case of the perfect cubic $P\,m\,\bar{3}\,m$ structure, and at least two O atoms and one Fe atom in the other phases, as illustrated in the bottom left diagram. 
    The locations of the Fe and O atoms are labelled in the $P\,m\,\bar{3}\,m$ section, and the hole effective masses, $m^{*}_{h}$ are shown below each panel for ease of reference. The full 3D data files are available in the SI \cite{Shenton2018}.
    }
    \label{fig:relaxed_ks-slices}
    \end{figure*}

Compared to the conduction bands, the topmost valence bands in Fig. \ref{fig:relaxed_pbands} show a far less dramatic difference in character across the five phases. All of the phases considered here have an O $p$ dominated $\mathrm{VB_{max}}$.
However, there are minor contributions from Fe $e_{g}$ states to the $\mathrm{VB_{max}}$, most notably in the $P\,m\,\bar{3}\,m$ phase.
To see more clearly the variation in the states at the $\mathrm{VB_{max}}$, we show in Fig. \ref{fig:relaxed_ks-slices} the cross-sections through KS orbitals corresponding to the topmost valence band. 
The figures show that, as $m^{*}_{h}$ decreases from $P\,n\,m\,a$, $R\,3\,c$, $R\,\bar{3}\,c$, $P\,4\,m\,m$ to $P\,m\,\bar{3}\,m$, the contribution from the Fe $e_g$ states increases. This observation indicates that the presence of Fe $e_{g}$ states at the $\mathrm{VB_{max}}$ plays a role in decreasing $m^{*}_{h}$.

\subsection{\label{subsec:geom_trans} Effects of structural transformations on effective masses \protect}

\begin{figure}
        \subfloat[Rotation \label{sfig:schem_rot}]{
            \includegraphics[ width=0.5\linewidth]{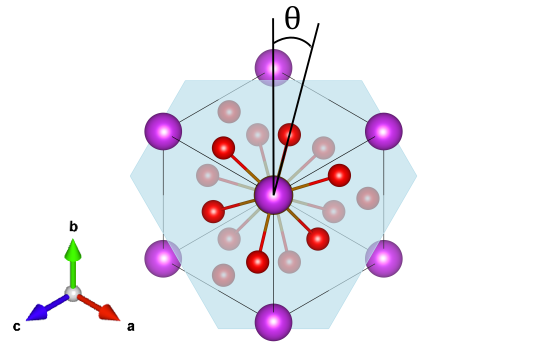}
        }
        \subfloat[Translation \label{sfig:schem_trans}]{
            \includegraphics[ width=0.5\linewidth]{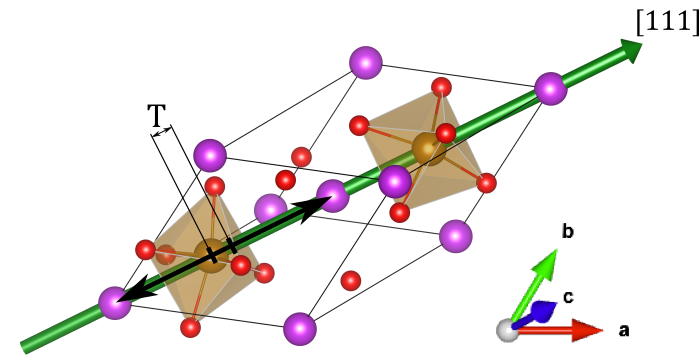}
        }
        \caption{Schematic of the \ce{FeO6} octahedral (a) rotations about, and (b) translations along the $[111]_{pc}$ direction present in the relaxed $R\,3\,c$ structure. In (a) we separate neighbouring octahedra along the $[111]_{pc}$ axis by a semi-transparent plane in order to highlight the out-of-phase nature of these rotations. The green arrow in (b) shows the $[111]_{pc}$ direction, and T is the displacement of the Fe atom from the mid-point between successive Bi atoms along this direction. Purple, ochre and red spheres represent Bi, Fe and O respectively. }
        \label{fig:geom_scheme}
    \end{figure}

\begin{figure}
        \centering
        \includegraphics[width=\linewidth]{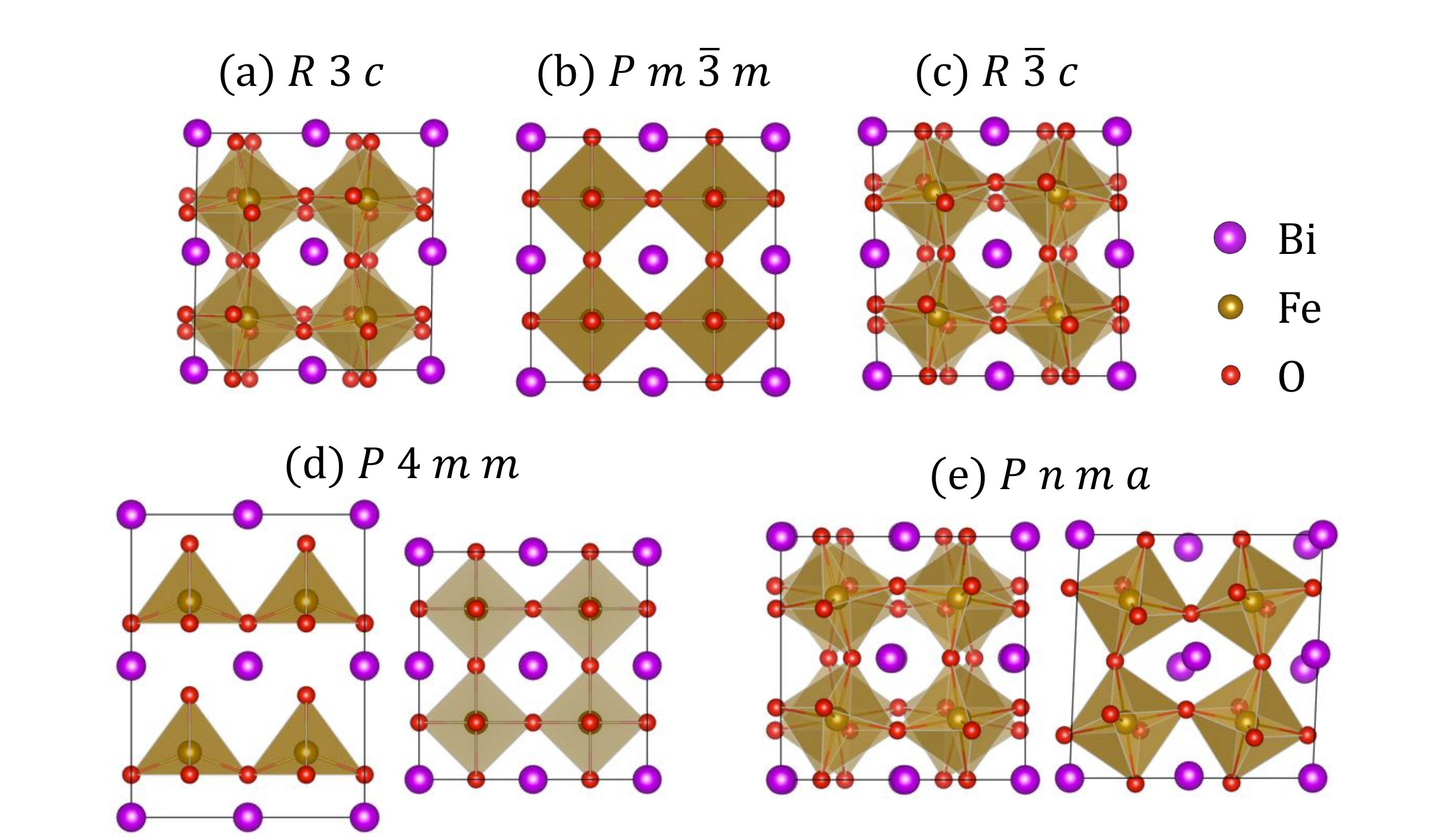}
        \caption{Crystal structures of five phases of BFO.
        For ease of comparison, all structures are presented in their cubic or pseudocubic settings. For the (a) $R\,3\,c$, (b) $R\,\bar{3}\,c$ and (c) $P\,m\,\bar{3}\,m$ structures, the three pseudocubic axes are equivalent. The (d) $P\,4\,m\,m$ and (e) $P\,n\,m\,a$ structures however have one nonequivalent axis. For these structures the left (right) figure has the nonequivalent axis parallel (perpendicular) to the page. Note that in the (d) $P\,4\,m\,m$ case, the elongated cell causes the \ce{FeO6} octahedra to break into square pyramidal \ce{FeO5} units.
            }
        \label{fig:relaxed_structs}
    \end{figure}

The different crystal structures of the BFO phases considered have been found to have a wide variation in the electronic structures and, in particular, in their charge carrier effective masses.
As such, a substantial yet complex relationship exists between crystal structure and effective mass in BFO.
In order to better understand this relationship, we now consider the geometric transformations that map the phase with the lowest effective masses, $P\,m\,\bar{3}\,m$, to the phase with the largest effective masses, $R\,3\,c$.
These transformations are: a) an antiphase rotation of the \ce{FeO6} octahedra about the $[111]_{pc}$ direction (i.e. $\mathrm{a^-a^-a^-}$ in Glazer's notation \cite{Glazer1972a}), and b) a translation of the \ce{FeO6} octahedra along the $[111]_{pc}$ direction. These transformations are illustrated in Fig. \ref{fig:geom_scheme}.
More generally, these and similar geometric transformations can describe the other BFO phases, portrayed in Fig. \ref{fig:relaxed_structs}, as follows:

\begin{itemize}
\item $P\,n\,m\,a$: an octahedral rotation out-of-phase along the $[100]_{pc}$ and $[010]_{pc}$ directions but in-phase along the $[001]_{pc}$ direction ($\mathrm{a^-a^-c^+}$); and alternating positive and negative translations in the $[010]_{pc}$ direction

\item $R\,\bar{3}\,c$: an out-of-phase rotation about the $[111]_{pc}$ axis ($\mathrm{a^-a^-a^-}$) with no translation

\item $P\, 4\, m\, m$: no rotation but with a substantial translation in the $[001]_{pc}$ direction
\end{itemize}

These transformations are accompanied by some strain to accommodate the change in the structure. We therefore consider the effects of strain, rotation and translation separately, particularly in relation to the ground state $R\,3\,c$ structure, which has shown remarkable structural flexibility in the context of heteroepitaxially grown BFO films \cite{Sando2014}.

    \subsubsection{\label{subsubsec:stretch} Strain\protect}

\begin{figure*}
        \centering
        \subfloat[\textbf{$m^{*}$} ]{\includegraphics[ width=0.35\linewidth]{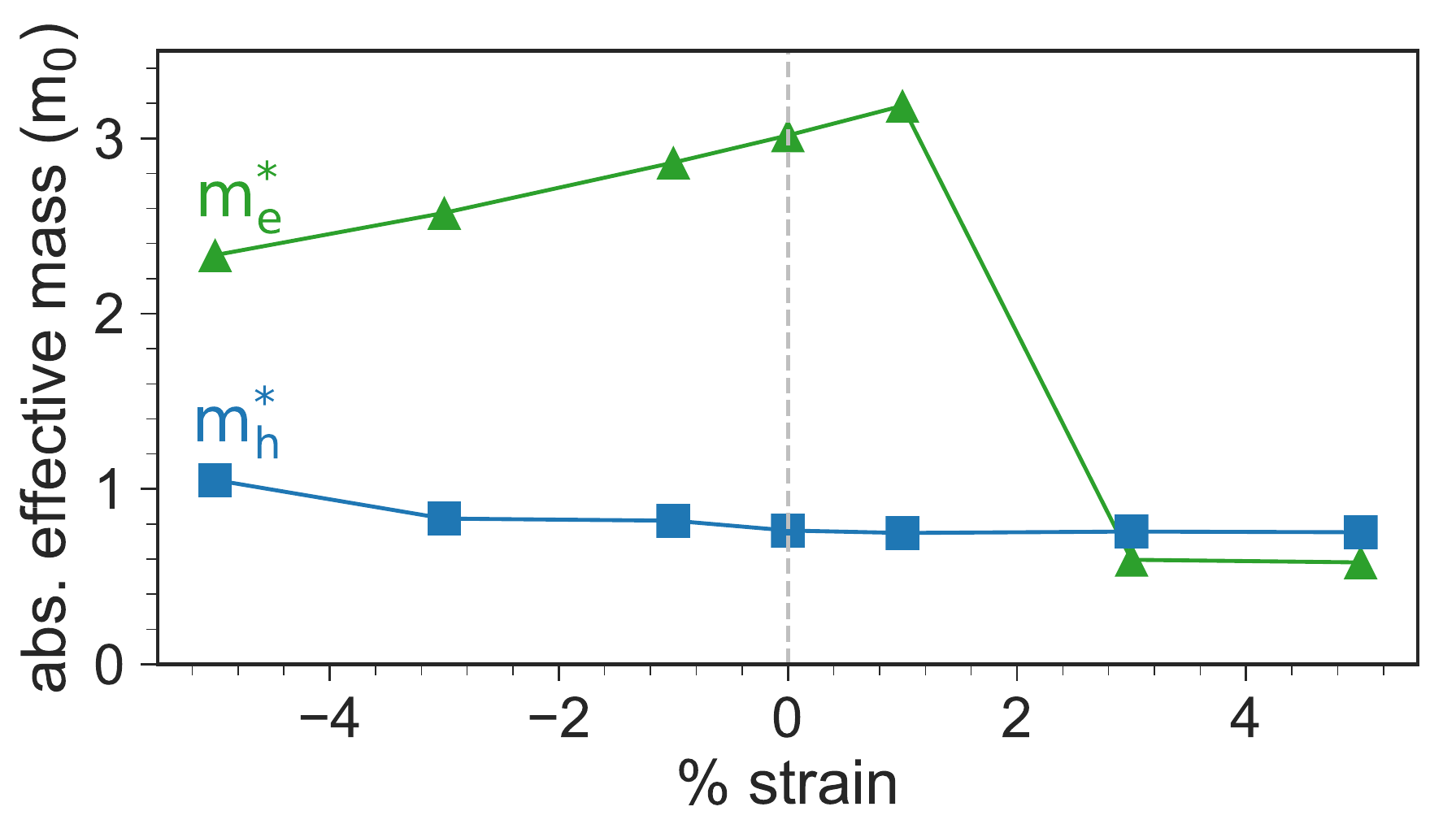}\label{sfig:stretching_mass}}
        
             \subfloat[\textbf{-5\% strain} ]{\includegraphics[width=0.3\linewidth]{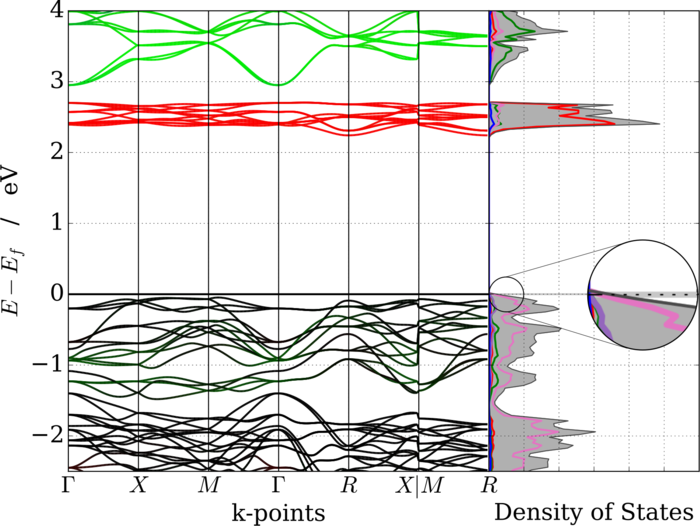}}\hspace{0.5em}%
             \subfloat[\textbf{ 0\% strain}]{\includegraphics[width=0.3\linewidth]{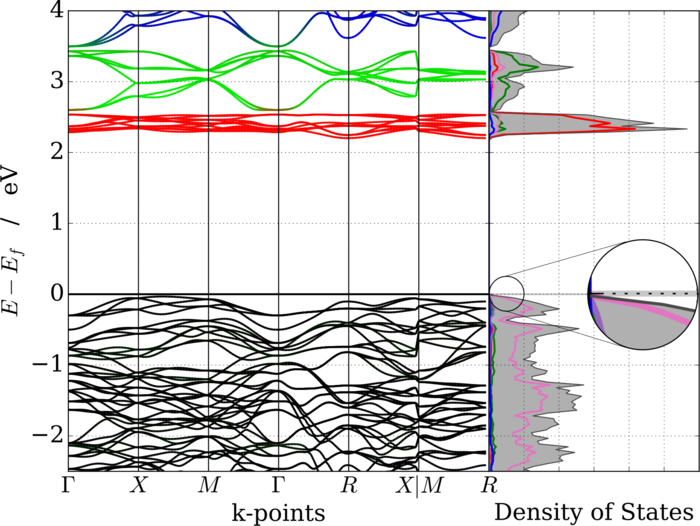}}\hspace{0.5em}%
             \subfloat[\textbf{+5\% strain}]{\includegraphics[width=0.3\linewidth]{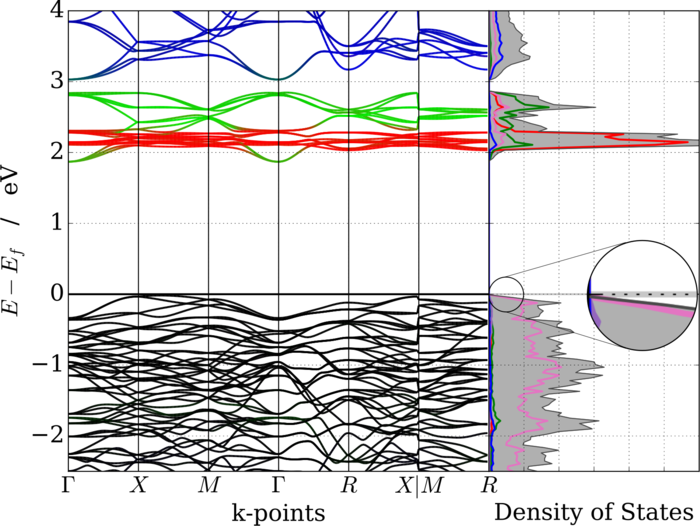}}
        
             \subfloat{\includegraphics[width=0.4\linewidth]{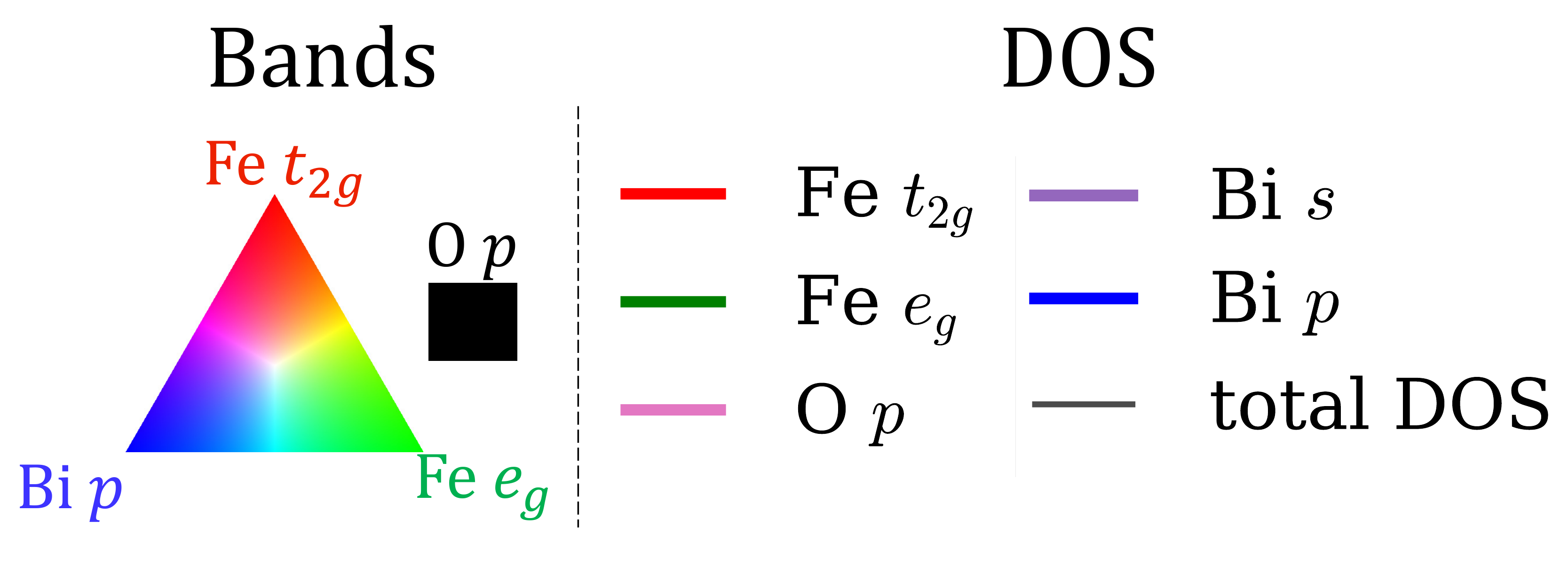}}
        \caption{Effects of uniform strain on the (a) $m^{*}_{e}$ (green triangles) and $m^{*}_{h}$ (blue squares), and (b-d) projected bands and density of states. Note that while the DOS axis has arbitrary units here, the scale is the same in each subfigure. The bands are coloured, at each $k$-point, based on wavefunction projections onto chosen orbitals. As indicated in the legend in panel, red, green, blue and black represent projections onto Fe $t_{2g}$, Fe $e_{g}$, Bi $p$ and O $p$ states respectively.}
        \label{fig:strain}
        \end{figure*}

    Various experimental works have found the $R\,3\,c$ phase of BFO to be stable over a large range of epitaxial strain: from $-$2.6\% to $+$1.2\% \cite{Rana2007, Infante2010, Sando2013}. 
    In addition, spontaneous polarisation in BFO has been found to be relatively insensitive to epitaxial strain values of up to $\pm$3\%. \cite{Ederer2005a, Ederer2005b}
    Epitaxial strain can also be used to stabilise other phases of BFO, such as the tetragonal $P\,4\,m\,m$ phase, as summarised by Sando \emph{et al.} \cite{Sando2014,Sando2016}.
    Since we are interested in the transformation from the $R\,3\,c$ to the $P\,m\,\bar{3}\,m$ phase, which involves an isotropic change in lattice constants, we consider the case of \emph{uniform}, rather than epitaxial strain. 
    We apply uniform strains between $-$5\% to $+$5\% to a BFO unit cell constrained to the $R\,3\,c$ symmetry. 
    Fig. \ref{sfig:stretching_mass} shows the effect of such strain on the charge carrier effective masses.
    
    The $m^{*}_{e}$ changes significantly, decreasing from 3.02 $m_0$ in the unstrained cell, to 2.33 $m_0$ under 5\% compressive strain.
    Under a tensile strain of between 1\% and 3\% we see an even larger decrease in $m^{*}_{e}$, from 3.19 $m_0$ to 0.60 $m_0$, corresponding to a change in the location and character of the $\mathrm{CB_{min}}$. Figs. \ref{fig:strain}(b-d) show the projected bands and DOS for the $-$5, 0 and $+$5\% strain cases.
   Similar figures for the other strain values are available in Fig. S2 of the SI.\cite{Shenton2018}
   From these figures a shift from an Fe $t_{2g}$ to an Fe $e_{g}$ dominated $\mathrm{CB_{min}}$ can be observed from 1\% to 3\% strain, which explains the sudden drop in $m^{*}_{e}$ for large tensile strain, since the $e_{g}$ bands are more dispersive than the $t_{2g}$ bands. We can understand the shift from a $t_{2g}$ to an $e_{g}$ dominated $\mathrm{CB_{min}}$ by considering the reduction in the splitting of the Fe $d$ orbitals, due to the octahedral environment, as the Fe-O bond lengths increase.
    The $m^{*}_{h}$ however, shows little dependence on strain. Over the range considered, $|m^{*}_{h}|$ is largest (1.051 $m_0$) at $-$5\% strain, and smallest (0.753 $m_0$) at $+$5\% strain. 
    For all strain values considered, the $\mathrm{VB_{max}}$ remains strongly dominated by O $p$ states as evident from Figs. \ref{fig:strain}(b-d). 
    Thus, the negligible changes in $m^{*}_{h}$ as the cell is strained are reflected by the minor changes in the character of the $\mathrm{VB_{max}}$.

    \subsubsection{\label{subsubsec:rotation}Rotation of the \ce{FeO6} octahedron\protect}

    \begin{figure}
        \centering
        \includegraphics[ width=\linewidth]{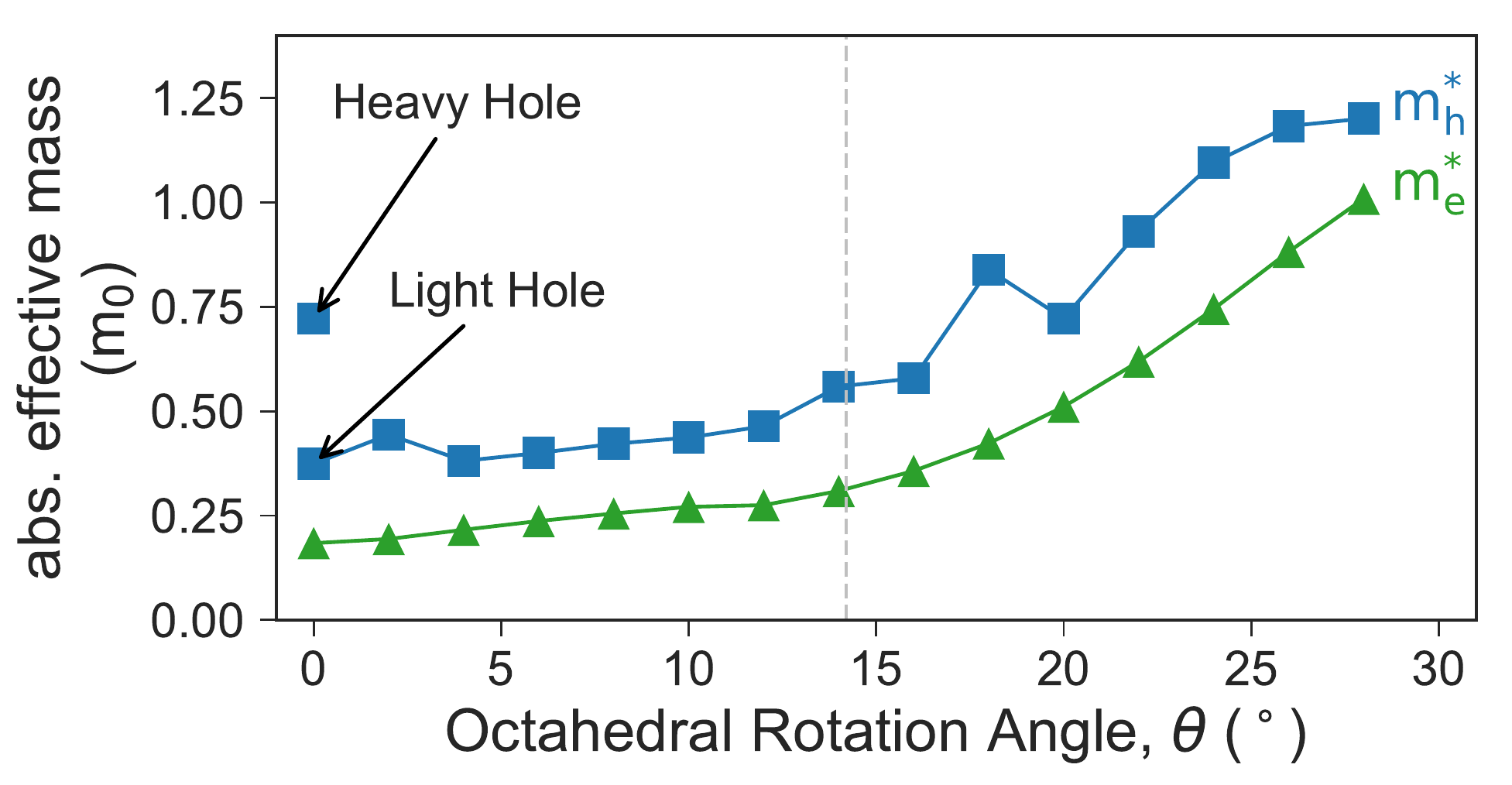}
        \caption{Absolute charge carrier effective mass versus octahedral rotation angle.
        The FeO$_6$ octahedra are rotated out-of-phase about the $[111]_{pc}$ axis.
        For each angle, $\theta$, the $m^{*}_{h}$ (blue squares) and $m^{*}_{e}$ (green triangles) were calculated.
        Notice that in the perfect cubic structure ($\theta=0^{\circ}$), there is a band degeneracy at the top of the VB - hence the heavy and light holes.
        This degeneracy breaks as soon as we have any octahedral rotation.
        The rotation angle of the $R\,3\,c$ PBE+U relaxed structure ($\theta\approx14.2^{\circ}$) is indicated with a vertical dashed line.
            }
        \label{fig:rotation_mass}
    \end{figure}

    \begin{figure*}
        \centering
            \subfloat[\textbf{$\theta=0^{\circ}$} ]{\includegraphics[width=0.31\linewidth]{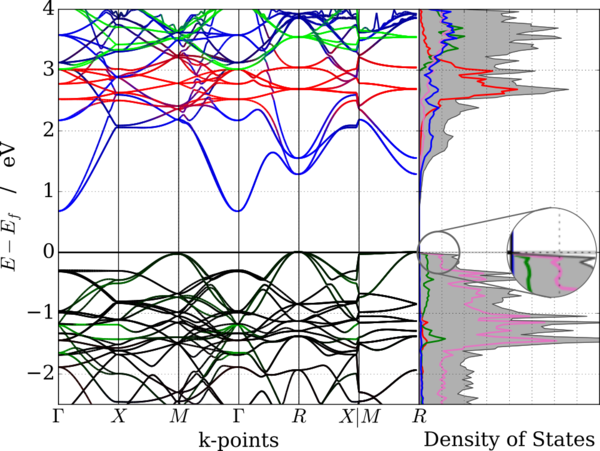}}\hspace{0.5em}%
            \subfloat[\textbf{$\theta=14^{\circ}$}]{\includegraphics[width=0.31\linewidth]{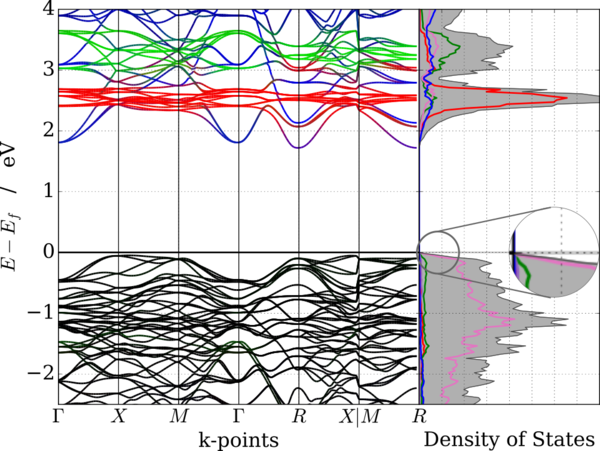}}\hspace{0.5em}%
            \subfloat[\textbf{$\theta=28^{\circ}$}]{\includegraphics[width=0.31\linewidth]{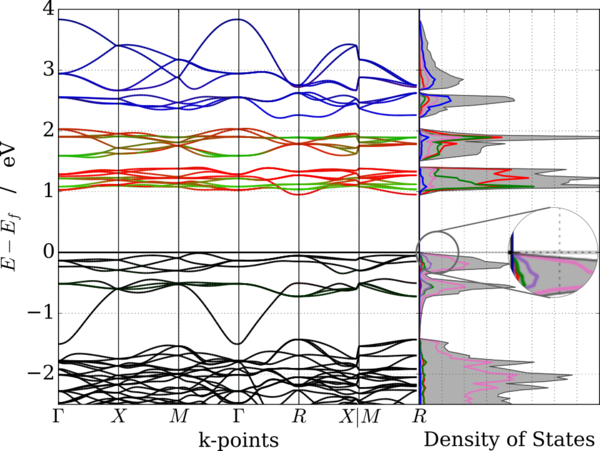}}

            \subfloat{\includegraphics[width=0.4\linewidth]{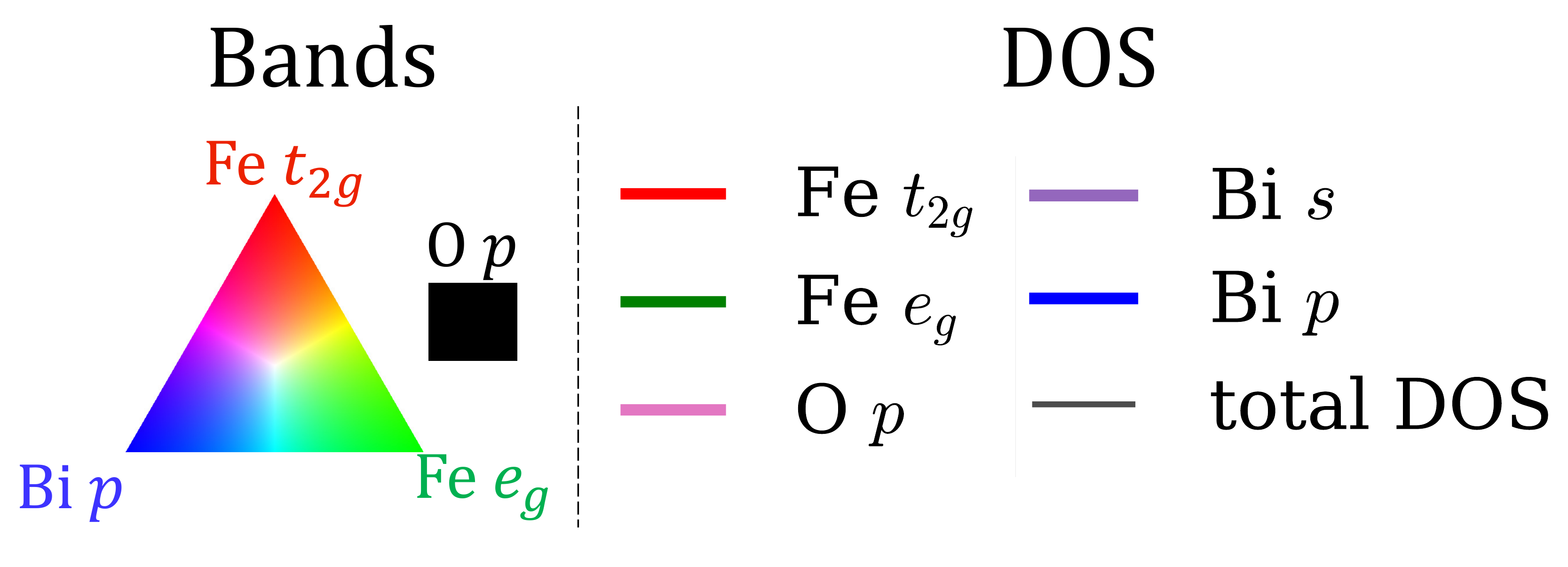}}
    \caption{Variation in projected bands and density of states as a function of out-of-phase \ce{FeO6} octahedral rotation about the $[111]_{pc}$ axis. Note that while the DOS axis has arbitrary units here, the scale is the same in each subfigure. The bands are coloured, at each $k$-point, based on wavefunction projections onto chosen orbitals. As indicated in the legend in panel, red, green, blue and black represent projections onto Fe $t_{2g}$, Fe $e_{g}$, Bi $p$ and O $p$ states respectively.}
    \label{fig:rotation_pbands}
    \end{figure*}

    To examine the effect of octahedral rotation on the electronic properties of BFO, we rotate the \ce{FeO6} octahedron from the perfect cubic perovskite geometry about the $[111]_{pc}$ axis in the out-of-phase manner shown in Fig. \ref{sfig:schem_rot}.
    In Fig. \ref{fig:rotation_mass} we plot the dependence of $m^{*}$ on the rotation angle, $\theta$.
    Both $m^{*}_{e}$ and $m^{*}_{h}$ increase with increasing $\theta$. Around $\theta=15^{\circ}$ we see a marked increase in the gradient of this relationship.

    To understand the trends in $m^*$ with $\theta$, we examine the electronic structures at $\mathrm{VB_{max}}$ and $\mathrm{CB_{min}}$ as a function of $\theta$.
    In Fig. \ref{fig:rotation_pbands} we plot the projected bands and DOS of BFO with the \ce{FeO6} octahedron rotated at
    $\theta=0^{\circ}$,
    $\theta=14^{\circ}$ and
    $\theta=28^{\circ}$.
    Similar figures for the full range of $\theta$ considered can be found in Fig. S6 of the SI.
    As $\theta$ increases up to $\theta=14^{\circ}$, the Bi $p$ bands move up in energy but we see little change in the character of the $\mathrm{CB_{min}}$, and correspondingly little change in $m^{*}_{e}$.
    Around 14$^{\circ}$ we see a shift in the location of the $\mathrm{CB_{min}}$ from $\Gamma$ to $R$.
    For $\theta> 14^{\circ}$, the character of the $\mathrm{CB_{min}}$ transitions from Bi $p$ to Fe $t_{2g}$, and then to a possible
    \footnote{Because we project the wavefunctions onto $d$ orbitals defined with respect to the \emph{global} Cartesian axes, the ability to resolve the difference between $t_{2g}$ and $e_g$ states diminishes as the local octahedral axes rotate relative to the global Cartesian axes. In other words, the `mixing' of the $t_{2g}$ and $e_{g}$ manifolds may be an artefact of the way the projections are done. To unambiguously resolve the relative contributions of  Fe $t_{2g}$ and $e_{g}$ states, one might perform a transformation from the global Cartesian basis to one that is local to each octahedron, as was done in Ref. \citenum{Mellan2015}. However, knowledge of the precise level of mixing between the Fe $t_{2g}$ and $e_{g}$ states goes beyond the requirements for the present study.}
     mix of Fe $e_{g}$ and $t_{2g}$ states for large $\theta$.
    The Bi $p$ to Fe $t_{2g}$ transition is associated with the large increase in $m^{*}_{e}$ with $\theta$ shown in Fig. \ref{fig:rotation_mass}.

    Within the topmost valence bands, we find that O $p$ states dominate for all values of $\theta$ considered. The states with minor contributions to the $\mathrm{VB_{max}}$ can be seen more clearly in the projected DOS. As $\theta$ increases, there is a decrease in the Fe $e_{g}$ contribution, and an increase in the Bi $s$ contribution to the top of the VB.
    We therefore associate the increase in $m^{*}_{h}$ with a decrease in Fe $e_{g}$ contribution and an increase in Bi $s$ contribution to the top of the valence band.
    Real-space plots of the KS orbitals as a function of $\theta$ can be found in Fig. S7 of the SI.
    We also observe that for $\theta = 0^{\circ}$, there is a band degeneracy at $\mathrm{VB_{max}}$, and hence we present $m^{*}_{h}$ values for both the heavy and light holes in Fig. \ref{fig:rotation_mass}. This degeneracy lifts as soon as we have any rotation.

    \subsubsection{\label{subsubsec:translations}Translation of the \ce{FeO6} octahedron \protect}

        \begin{figure}
            \subfloat[$\theta=0^{\circ}$ \label{sfig:trans0}]{
                \includegraphics[ width=\linewidth]{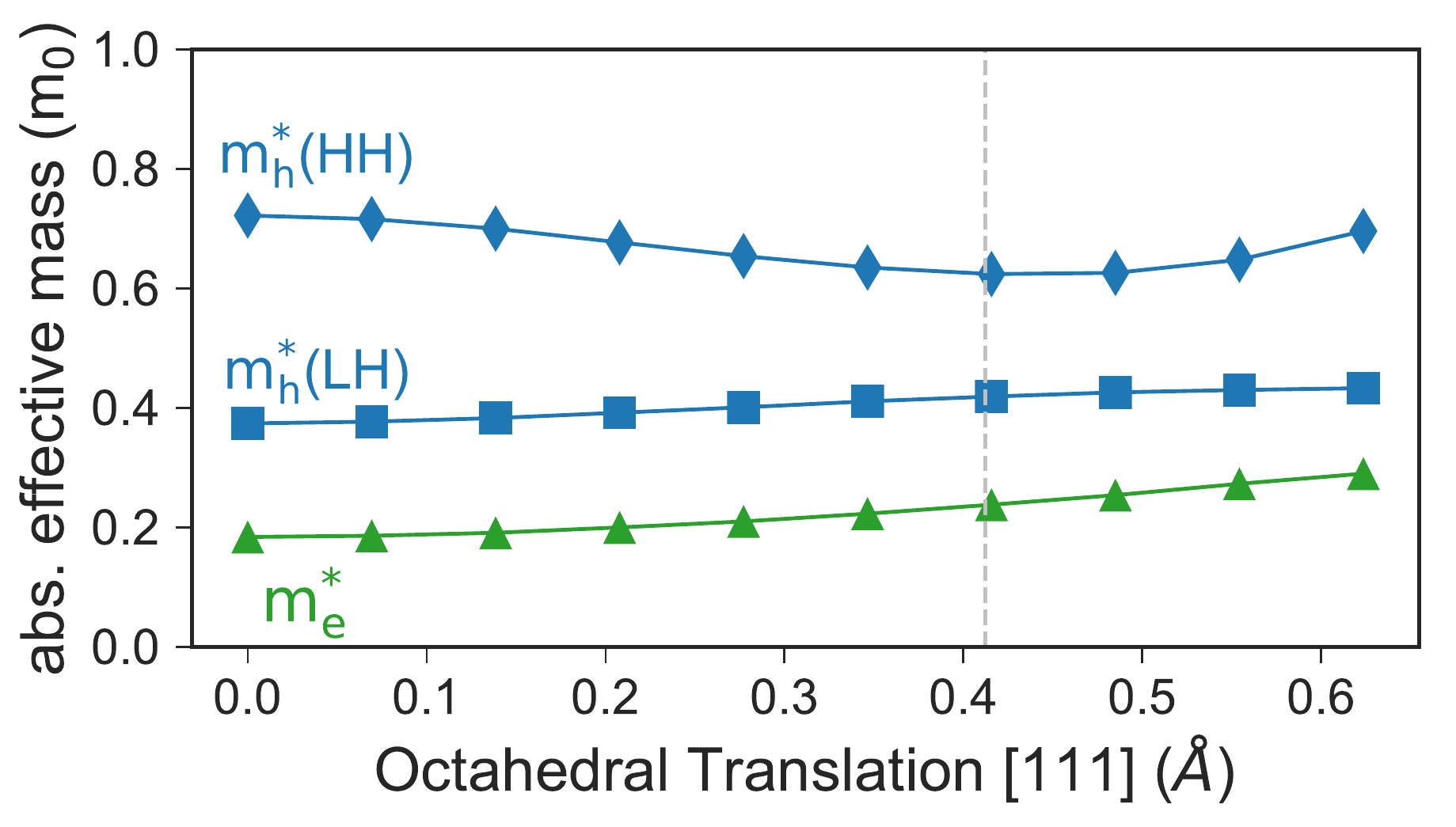}
            }\hfill
            \subfloat[$\theta=14^{\circ}$ \label{sfig:trans14}]{
                \includegraphics[ width=\linewidth]{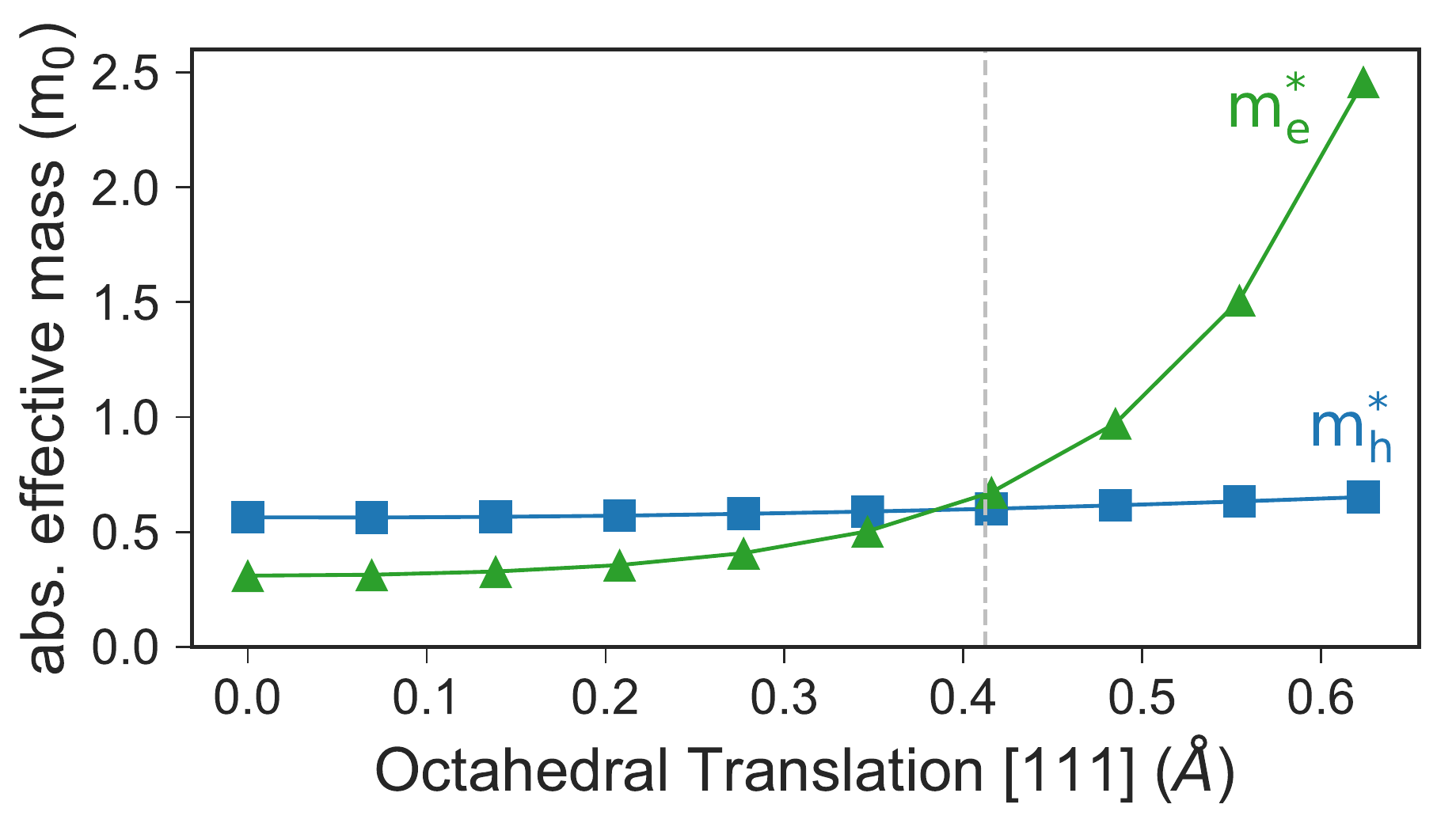}
            }
            \caption{Absolute charge carrier effective mass versus $[111]_{pc}$ octahedral translation.
             In (a) we translate the perfect FeO$_6$ octahedron ($\theta=0^{\circ}$), while in (b) we translate a distorted octahedron ($\theta=14^{\circ}$).
             The angle, $\theta$, is as defined in section \ref{subsubsec:rotation}. 
             The $m^{*}_{e}$ and $m^{*}_{h}$  are represented by green triangles and blue squares respectively. In (a), the band degeneracy at $\mathrm{VB_{max}}$ gives rise to a heavy hole, HH (blue diamonds), in addition to the light hole, LH (blue squares).
             The vertical lines in each subfigure correspond to the PBE+U relaxed $R\,3\,c$ Fe translation along $[111]_{pc}$ ($\approx 0.412$ \AA). }
            \label{fig:translation_mass}
        \end{figure}

        \begin{figure*}
            \centering
             \subfloat[$T=0.00$ \AA, $\theta=0^{\circ}$]{\includegraphics[width=0.30\linewidth]{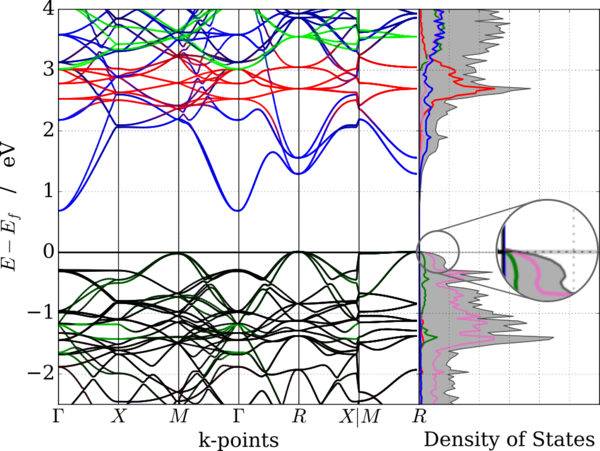}}\hspace{0.5em}%
             \subfloat[$T=0.42$ \AA, $\theta=0^{\circ}$]{\includegraphics[width=0.30\linewidth]{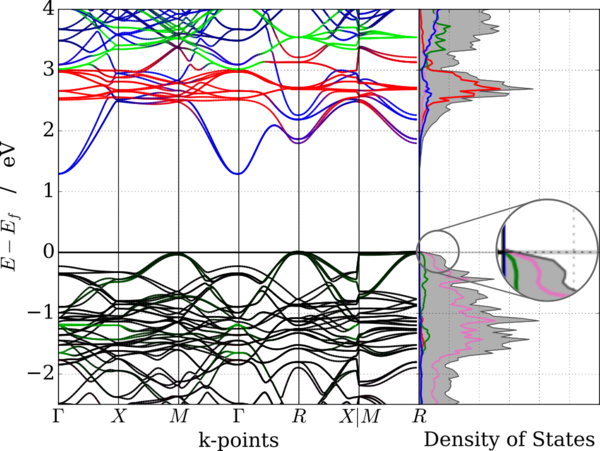}}\hspace{0.5em}%
             \subfloat[$T=0.62$ \AA, $\theta=0^{\circ}$]{\includegraphics[width=0.30\linewidth]{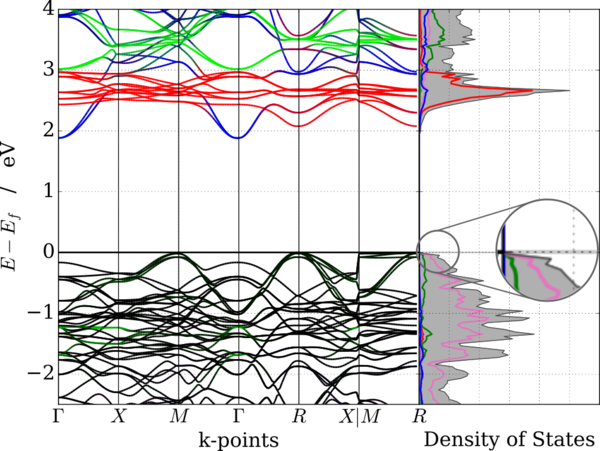}}

             \subfloat[$T=0.00$ \AA, $\theta=14^{\circ}$]{\includegraphics[width=0.30\linewidth]{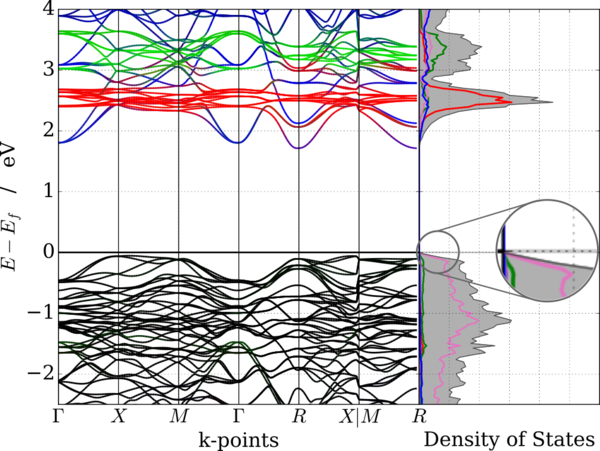}}\hspace{0.5em}%
             \subfloat[$T=0.42$ \AA, $\theta=14^{\circ}$]{\includegraphics[width=0.30\linewidth]{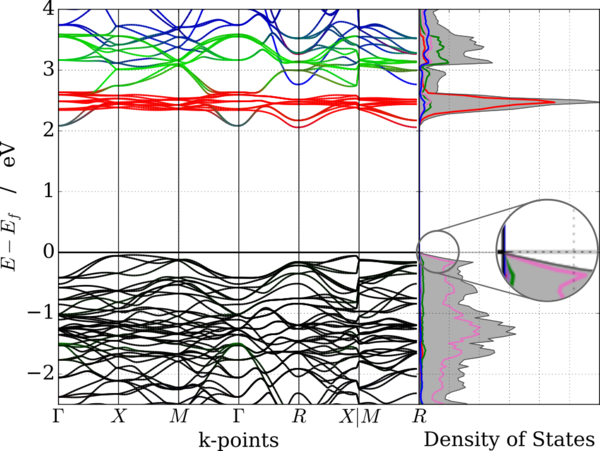}}\hspace{0.5em}%
             \subfloat[$T=0.62$ \AA, $\theta=14^{\circ}$]{\includegraphics[width=0.30\linewidth]{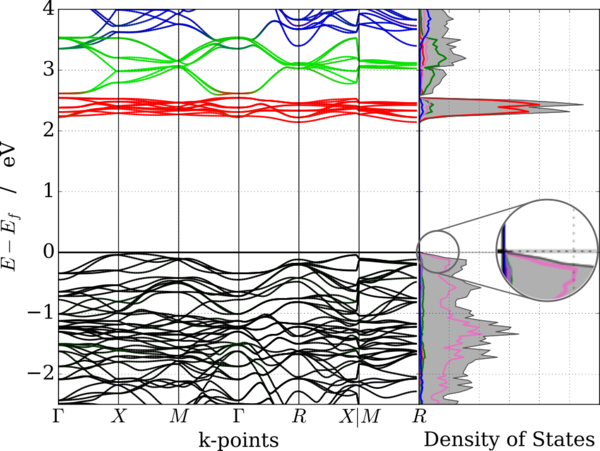}}

             \subfloat{\includegraphics[width=0.4\linewidth]{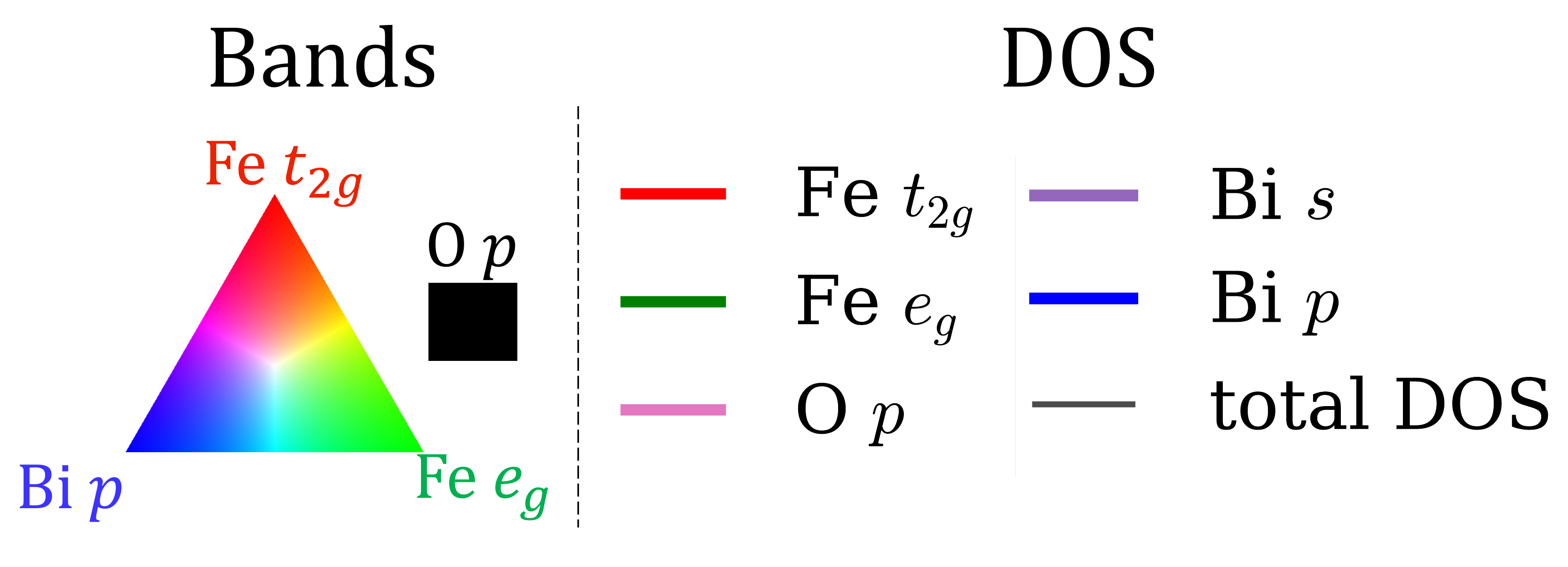}}

            \caption{Variation in projected bands and density of states with  FeO$_6$ octahedral translation along the $[111]_{pc}$ direction.
             $\theta=0^{\circ}$ and $\theta=14^{\circ}$ indicates the translation of the perfect (a-c) and distorted octahedra (d-f) respectively and T gives the translation magnitude.
             The bands are coloured based on projections onto chosen orbitals. As indicated in the legend, red, green, blue and black represent projections onto Fe $t_{2g}$, Fe $e_{g}$, Bi $p$ and O $p$ states respectively.}
            \label{fig:translation_pbands}
        \end{figure*}

    We consider, independently, translations of distorted (rotated $\theta = 14^{\circ}$ about the $[111]_{pc}$ axis) and perfect ($\theta = 0^{\circ}$) FeO$_6$ octahedra along the $[111]_{pc}$ direction. The former translation is as illustrated in Fig. \ref{sfig:schem_trans}. 
    We find a marked difference between the two cases, as shown in Fig. \ref{fig:translation_mass} where we plot the dependence of $m^{*}_{e}$ and $m^{*}_{h}$ on the magnitude of the translation, T.
    Translating the perfect octahedron (Fig. \ref{sfig:trans0}) has negligible effects on both $m^{*}_{e}$ and $m^{*}_{h}$.
    Translating the distorted octahedron (Fig. \ref{sfig:trans14}) also has little effect on $m^{*}_{h}$; however, there is a large effect on $m^{*}_{e}$.

    Fig. \ref{fig:translation_pbands} shows changes in the projected bands and DOS as the octahedra are translated. Once again we observe that changes in $m^{*}$ are correlated with changes in the chemical character of the band extrema.
    Figs. \ref{fig:translation_pbands} (a-c) confirm that the character at the $\mathrm{CB_{min}}$ of the structure with the perfect octahedron does not change within the range of T considered.
    Figs. \ref{fig:translation_pbands} (d-f), however, show a shift from Bi $p$ to Fe $t_{2g}$ character in the case of the $14^{\circ}$ rotated octahedron, corresponding to the substantial increase in $m^{*}_{e}$.
    Conversely, there is no change in character at the $\mathrm{VB_{max}}$, and also little change in $m^{*}_{h}$ with T, for both the perfect and the rotated octahedra.
    The $\mathrm{VB_{max}}$ degeneracy at $\theta=0^\circ$, that was lifted with octahedral rotations in Sec. \ref{subsubsec:rotation}, does not vanish in the case of octahedral translations, as indicated in Figs. \ref{fig:translation_pbands}(a-c).


\section{\label{sec:discuss}Discussion\protect}

In our investigations of the electronic properties in BFO, a clear trend between the orbital characters at the band edges and the resulting $m^*$ emerges. 
We find that presence of Fe $e_g$ states at the $\mathrm{VB_{max}}$ and the Bi $p$ states at the $\mathrm{CB_{min}}$ lead to relatively low $m^*$. 
At the $\mathrm{VB_{max}}$, hybridisation between the Fe $e_g$ and O 2$p$ states leads to a more dispersive band and hence a lower $m^{*}_{h}$. 
In the conduction bands, the Bi $p$ states are more dispersive than the Fe $t_{2g}$, and hence structures that have a $\mathrm{CB_{min}}$ that is dominated by Bi $p$ states have a lower $m^{*}_{e}$ than those whose $\mathrm{CB_{min}}$ is dominated by Fe $t_{2g}$ states.

We have demonstrated that the changes in the band edge characters and hence $m^*$ can be effected through geometric changes to the crystal structure. Extreme tensile strain, suppression of the octahedral rotation and suppression of the octahedral translation yield a much reduced $m^*$ in BFO. This trend explains the variation in $m^*$ across the BFO phases. The high $m^*$ values of the $R\,3\,c$ and $P\,n\,m\,a$ phases can be attributed to the presence of octahedral rotation and translation in these phases. 
When we compare the $R\,3\,c$ and $R\,\bar{3}\,c$ structures, which contain similarly rotated octahedra, we find that the phase with the translated octahedron ($R\,3\,c$) has larger $m^{*}$ values compared with the untranslated phase ($R\,\bar{3}\,c$).
Interestingly, we find that when the rotation is completely suppressed ($\theta=0$), translation of the octahedron (within the range of study) has little effect on $m^*$. Hence the $P\, 4\, m\, m$ structure, which has translation without rotation, also has lower $m^*$ values than those of $R\,3\,c$. The $P\,m\,\bar{3}\,m$ structure has the lowest $m^*$ values of all the structures considered since it has neither octahedral rotation or translation.

For applications in light harvesting, materials with low $m^*$ values are required to allow the charge carriers to be transported to the electrode before the carriers recombine. Appreciable photovoltaic responses have already been recorded in BFO in the ground-state $R\,3\,c$ phase \cite{Choi2009,Yang2009b,Ji2010,Paillard2016,Wu2014}. Our study indicates that its crystal structure may be modified to optimise $m^*$. Particularly in thin film heteroepitaxial growth, some degree of control over the octahedral tilt angle and tilt pattern has been demonstrated experimentally in BFO and other oxide systems \cite{Borisevich2010,Kim2013b,He2010,Kan2016,Rondinelli2012}. 
While the high temperature $P\,m\,\bar{3}\,m$ phase of BFO has the required geometric features to result in low $m^{*}$, it lacks the asymmetric potential required for the bulk photovoltaic effect.
We find that the tetragonal $P\,4\,m\,m$ phase, which is experimentally accessible under compressive epitaxial strain, has both a large spontaneous polarisation and relatively low $m^{*}$.
However, the $P\, 4\, m\, m$ phase has been reported to have a larger optical gap than the $R\,3\,c$ phase, despite the smaller electronic gap, which would adversely affect the efficiency of the device \cite{Chen2010}. Further work is required to find a balance between these competing effects. Besides geometry manipulation, the band edge character can also be modified through chemical doping \cite{Wang2015w, Rong2015}, which may be an interesting subject for future work.

Another avenue for future work would be to consider the effects of the crystal structure on charge carrier relaxation times, since the electron and hole \emph{mobilities} of a given phase are, of course, not only determined by the effective mass of the charge carriers, but also by their relaxation times. These relaxation times depend on electron--phonon scattering \cite{Bardeen1950} and are thus expected to also be sensitive to changes in crystal structure.

As discussed above, the `curvature' definition of $m^*$ adopted here is one that emphasises the importance of the band edges in conduction. However, in the case of materials that have fairly flat bands with multiple valleys of similar energy, the band curvature at these local minima (/maxima) may also influence $m^*$. In Fig. \ref{fig:relaxed_pbands} we see that, particularly in the $R\,3\,c$ and $P\,n\,m\,a$ phases, the valence bands of BFO, do indeed exhibit several local maxima of similar energy.
In order to explore the effect of such multiple valleys, we calculate the `conductivity mass' as defined in Eq. 12 of Ref. \citenum{Madsen2006} and implemented in \textsc{BoltzTraP2} \cite{BoltzTraP2} for the five phases of BFO considered above. The electron conductivity masses are found to be in good quantitative agreement with `curvature masses' listed in Table \ref{tab:relaxed}. The hole conductivity masses, however, are generally larger than the hole curvature masses listed in Table \ref{tab:relaxed}, consistent with the presence of multiple local maxima. More importantly for the purposes of the present work, the trends in both $m_e^*$ and $m_h^*$ generally agree between the two effective mass definitions. An exception to this is the hole effective mass of the $P\,n\,m\,a$ phase. Whilst $P\,n\,m\,a$ has the largest hole curvature mass, its conductivity mass is slightly smaller than those of the $R\,\bar{3}\,c$ and $R\,3\,c$ phases. Despite this, given the otherwise good agreement in the $m^*$ trends across these BFO phases, our tests indicate that the central conclusions drawn in this paper are not dependent on the choice of effective mass definition.
The conductivity mass tensors for the five BFO phases considered are also available in the SI \cite{Shenton2018}.

In addition to the intrinsic factors considered above, a number of extrinsic factors may also play a role in determining the $m^*$ in a given device. Factors such as temperature, charge carrier doping concentrations, impurities, grain boundaries, and FE domain walls could all play important roles in determining $m^*$.
However, since our aim is to compare $m^{*}$ between the BFO phases, and to suggest approaches to improve $m^{*}$ further, we only consider \emph{differences} in $m^{*}$.
That is to say, while factors such as temperature are likely to affect $m^{*}$, we have made the assumption in this work that our 0 K results are a necessary precursor to including high temperature effects.

The effect of spin-orbit coupling (SOC) has been found to be significant in describing certain properties of BFO, such as its weak (Dzyaloshinskii-Moriya) ferromagnetism \cite{Ederer2005}.
We find a noticeable shift in the energy of the electronic bands and a small, Rashba-like shift in the location of the band edges. For example, the inclusion of SOC for the $R\,3\,c$ structure shifts the location of the $\mathrm{CB_{min}}$ slightly off the high-symmetry $R$ point. Such a shift might improve the PV prospects of BFO by inhibiting charge carrier recombination. This Rashba effect, together with the effects of SOC on the $m^{*}$ anisotropy, while beyond the scope of the present study, would be an interesting avenue for further investigation.

Despite the changes to the positions of the bands, the band curvatures around the Fermi level (and hence $m^{*}$) are largely unaffected by SOC. For the $R\,3\,c$ structure, we calculate the hole effective mass to be  -0.780 $m_0$ with SOC and -0.669 $m_0$ without SOC. Similarly, the electron effective mass changed from 2.89 $m_0$ with SOC, to 3.06 $m_0$ without SOC. Note that in these comparisons we have kept constant the direction along which we evaluate $m^{*}$: $[-0.28366, 1.00000, -0.17603]$ for the hole mass and $[1,1,1]$ for the electron mass. These correspond to the principle directions that yield the lowest $m^{*}$ in the non-SOC case; i.e. those that were used in Table \ref{tab:relaxed} for $R\,3\,c$. 
Such differences are roughly an order of magnitude smaller than the differences in $m^{*}$ that we found by comparing the various phases of BFO, and hence we neglect the effects of SOC throughout this work.
Further tests of the effect of SOC on the character of bands as a function of octahedral rotations are available in Fig. S6 \cite{Shenton2018}.


\section{\label{sec:conclusions}Conclusions\protect}

An important and often limiting factor in perovskite-based photovoltaics is their low charge carrier mobility. In this work we have employed first-principles methods to compare the charge carrier effective masses of both FE and non-FE phases of BFO and explore how the effective masses are influenced by changes in the crystal geometry. 
We find that the ground-state, FE $R\,3\,c$ phase has relatively large $m^{*}$ values compared to those of some non-FE phases. However, these non-FE phases lack the mechanism required for charge separation via the bulk photovoltaic effect. 
We therefore investigate the $m^{*}$-determining factors in BFO, with the aim of uncovering a FE phase with reduced $m^*$ values. 
We discover that the differences in the $m^*$ values are related to the orbital character at the band edges. 
BFO structures with Bi $p$ or Fe $e_g$ states, instead of Fe $t_{2g}$ states, at the $\mathrm{CB_{min}}$ have lower $m^{*}_{e}$ due to the localised nature of the $t_{2g}$ bands. Structures with hybridised Fe $e_{g}$ - O $p$ states at the $\mathrm{VB_{max}}$, rather than pure O $p$ states, have lower $m^{*}_{h}$ since the hybridisation leads to a more delocalised state.
The change in the states at the $\mathrm{CB_{min}}$ leading to decreased $m^{*}_{e}$ can be achieved with: (i) high tensile uniform strain, (ii) reduction of $[111]_{pc}$  translations of distorted Fe-O octahedra, and (iii) a reduction in octahedral rotation about the $[111]_{pc}$ axis. The Fe $e_g$ states emerges at the $\mathrm{VB_{max}}$ only with a reduction in octahedral rotations about $[111]_{pc}$. 
These findings explain the reasons why the $P\,m\,\bar{3}\,m$ phase, which has unrotated and untranslated octahedra, has the lowest $m^{*}$ values of the BFO phases we have considered in this work. However, since we seek a FE phase and we find that $m^{*}$ changes little with translation when the octahedral rotation about $[111]_{pc}$ is suppressed, then a structure resembling the tetragonal $P\,4\,m\,m$ phase would provide both ferroelectricity and reduced $m^{*}$.
Our results demonstrate that manipulation of crystal geometry, which is easily achieved with advanced growth techniques, is a viable avenue to tune the electronic properties of materials particularly at the band edges.



 \begin{acknowledgments}
The authors acknowledge the use of the UCL Legion (Legion@UCL) and Grace (Grace@UCL) High Performance Computing Facilities, and associated support services, in the completion of this work.
\end{acknowledgments}


\begin{thebibliography}{57}%
\makeatletter
\providecommand \@ifxundefined [1]{%
 \@ifx{#1\undefined}
}%
\providecommand \@ifnum [1]{%
 \ifnum #1\expandafter \@firstoftwo
 \else \expandafter \@secondoftwo
 \fi
}%
\providecommand \@ifx [1]{%
 \ifx #1\expandafter \@firstoftwo
 \else \expandafter \@secondoftwo
 \fi
}%
\providecommand \natexlab [1]{#1}%
\providecommand \enquote  [1]{``#1''}%
\providecommand \bibnamefont  [1]{#1}%
\providecommand \bibfnamefont [1]{#1}%
\providecommand \citenamefont [1]{#1}%
\providecommand \href@noop [0]{\@secondoftwo}%
\providecommand \href [0]{\begingroup \@sanitize@url \@href}%
\providecommand \@href[1]{\@@startlink{#1}\@@href}%
\providecommand \@@href[1]{\endgroup#1\@@endlink}%
\providecommand \@sanitize@url [0]{\catcode `\\12\catcode `\$12\catcode
  `\&12\catcode `\#12\catcode `\^12\catcode `\_12\catcode `\%12\relax}%
\providecommand \@@startlink[1]{}%
\providecommand \@@endlink[0]{}%
\providecommand \url  [0]{\begingroup\@sanitize@url \@url }%
\providecommand \@url [1]{\endgroup\@href {#1}{\urlprefix }}%
\providecommand \urlprefix  [0]{URL }%
\providecommand \Eprint [0]{\href }%
\providecommand \doibase [0]{http://dx.doi.org/}%
\providecommand \selectlanguage [0]{\@gobble}%
\providecommand \bibinfo  [0]{\@secondoftwo}%
\providecommand \bibfield  [0]{\@secondoftwo}%
\providecommand \translation [1]{[#1]}%
\providecommand \BibitemOpen [0]{}%
\providecommand \bibitemStop [0]{}%
\providecommand \bibitemNoStop [0]{.\EOS\space}%
\providecommand \EOS [0]{\spacefactor3000\relax}%
\providecommand \BibitemShut  [1]{\csname bibitem#1\endcsname}%
\let\auto@bib@innerbib\@empty
\bibitem [{\citenamefont {Chynoweth}(1956)}]{Chynoweth1956}%
  \BibitemOpen
  \bibfield  {author} {\bibinfo {author} {\bibfnamefont {A.~G.}\ \bibnamefont
  {Chynoweth}},\ }\href {\doibase 10.1103/PhysRev.102.705} {\bibfield
  {journal} {\bibinfo  {journal} {Physical Review}\ }\textbf {\bibinfo {volume}
  {102}},\ \bibinfo {pages} {705} (\bibinfo {year} {1956})}\BibitemShut
  {NoStop}%
\bibitem [{\citenamefont {Choi}\ \emph {et~al.}(2009)\citenamefont {Choi},
  \citenamefont {Lee}, \citenamefont {Choi}, \citenamefont {Kiryukhin},\ and\
  \citenamefont {Cheong}}]{Choi2009}%
  \BibitemOpen
  \bibfield  {author} {\bibinfo {author} {\bibfnamefont {T.}~\bibnamefont
  {Choi}}, \bibinfo {author} {\bibfnamefont {S.}~\bibnamefont {Lee}}, \bibinfo
  {author} {\bibfnamefont {Y.~J.}\ \bibnamefont {Choi}}, \bibinfo {author}
  {\bibfnamefont {V.}~\bibnamefont {Kiryukhin}}, \ and\ \bibinfo {author}
  {\bibfnamefont {S.-W.}\ \bibnamefont {Cheong}},\ }\href {\doibase
  10.1126/science.1168636} {\bibfield  {journal} {\bibinfo  {journal}
  {Science}\ }\textbf {\bibinfo {volume} {324}},\ \bibinfo {pages} {63}
  (\bibinfo {year} {2009})}\BibitemShut {NoStop}%
\bibitem [{\citenamefont {Fridkin}(2001)}]{Fridkin2001}%
  \BibitemOpen
  \bibfield  {author} {\bibinfo {author} {\bibfnamefont {V.~M.}\ \bibnamefont
  {Fridkin}},\ }\href {\doibase 10.1134/1.1387133} {\bibfield  {journal}
  {\bibinfo  {journal} {Crystallography Reports}\ }\textbf {\bibinfo {volume}
  {46}},\ \bibinfo {pages} {654} (\bibinfo {year} {2001})}\BibitemShut
  {NoStop}%
\bibitem [{\citenamefont {Lebeugle}\ \emph {et~al.}(2007)\citenamefont
  {Lebeugle}, \citenamefont {Colson}, \citenamefont {Forget},\ and\
  \citenamefont {Viret}}]{Lebeugle2007}%
  \BibitemOpen
  \bibfield  {author} {\bibinfo {author} {\bibfnamefont {D.}~\bibnamefont
  {Lebeugle}}, \bibinfo {author} {\bibfnamefont {D.}~\bibnamefont {Colson}},
  \bibinfo {author} {\bibfnamefont {A.}~\bibnamefont {Forget}}, \ and\ \bibinfo
  {author} {\bibfnamefont {M.}~\bibnamefont {Viret}},\ }\href {\doibase
  10.1063/1.2753390} {\bibfield  {journal} {\bibinfo  {journal} {Applied
  Physics Letters}\ }\textbf {\bibinfo {volume} {91}},\ \bibinfo {pages}
  {022907} (\bibinfo {year} {2007})}\BibitemShut {NoStop}%
\bibitem [{\citenamefont {Shvartsman}\ \emph {et~al.}(2007)\citenamefont
  {Shvartsman}, \citenamefont {Kleemann}, \citenamefont {Haumont},\ and\
  \citenamefont {Kreisel}}]{Shvartsman2007}%
  \BibitemOpen
  \bibfield  {author} {\bibinfo {author} {\bibfnamefont {V.~V.}\ \bibnamefont
  {Shvartsman}}, \bibinfo {author} {\bibfnamefont {W.}~\bibnamefont
  {Kleemann}}, \bibinfo {author} {\bibfnamefont {R.}~\bibnamefont {Haumont}}, \
  and\ \bibinfo {author} {\bibfnamefont {J.}~\bibnamefont {Kreisel}},\ }\href
  {\doibase 10.1063/1.2731312} {\bibfield  {journal} {\bibinfo  {journal}
  {Applied Physics Letters}\ }\textbf {\bibinfo {volume} {90}},\ \bibinfo
  {pages} {172115} (\bibinfo {year} {2007})}\BibitemShut {NoStop}%
\bibitem [{\citenamefont {Yuan}\ \emph {et~al.}(2014)\citenamefont {Yuan},
  \citenamefont {Xiao}, \citenamefont {Yang},\ and\ \citenamefont
  {Huang}}]{Yuan2014}%
  \BibitemOpen
  \bibfield  {author} {\bibinfo {author} {\bibfnamefont {Y.}~\bibnamefont
  {Yuan}}, \bibinfo {author} {\bibfnamefont {Z.}~\bibnamefont {Xiao}}, \bibinfo
  {author} {\bibfnamefont {B.}~\bibnamefont {Yang}}, \ and\ \bibinfo {author}
  {\bibfnamefont {J.}~\bibnamefont {Huang}},\ }\href {\doibase
  10.1039/C3TA14188H} {\bibfield  {journal} {\bibinfo  {journal} {Journal of
  Materials Chemistry A}\ }\textbf {\bibinfo {volume} {2}},\ \bibinfo {pages}
  {6027} (\bibinfo {year} {2014})}\BibitemShut {NoStop}%
\bibitem [{\citenamefont {Butler}\ \emph {et~al.}(2015)\citenamefont {Butler},
  \citenamefont {Frost},\ and\ \citenamefont {Walsh}}]{Butler2015}%
  \BibitemOpen
  \bibfield  {author} {\bibinfo {author} {\bibfnamefont {K.~T.}\ \bibnamefont
  {Butler}}, \bibinfo {author} {\bibfnamefont {J.~M.}\ \bibnamefont {Frost}}, \
  and\ \bibinfo {author} {\bibfnamefont {A.}~\bibnamefont {Walsh}},\ }\href
  {\doibase 10.1039/C4EE03523B} {\bibfield  {journal} {\bibinfo  {journal}
  {Energy \& Environmental Science}\ }\textbf {\bibinfo {volume} {8}},\
  \bibinfo {pages} {838} (\bibinfo {year} {2015})}\BibitemShut {NoStop}%
\bibitem [{\citenamefont {Huang}(2010)}]{Huang2010}%
  \BibitemOpen
  \bibfield  {author} {\bibinfo {author} {\bibfnamefont {H.}~\bibnamefont
  {Huang}},\ }\href {\doibase 10.1038/nphoton.2010.15} {\bibfield  {journal}
  {\bibinfo  {journal} {Nature Photonics}\ }\textbf {\bibinfo {volume} {4}},\
  \bibinfo {pages} {134} (\bibinfo {year} {2010})}\BibitemShut {NoStop}%
\bibitem [{\citenamefont {Sun}\ \emph {et~al.}(2007)\citenamefont {Sun},
  \citenamefont {Thompson},\ and\ \citenamefont {Nishida}}]{Sun2007}%
  \BibitemOpen
  \bibfield  {author} {\bibinfo {author} {\bibfnamefont {Y.}~\bibnamefont
  {Sun}}, \bibinfo {author} {\bibfnamefont {S.~E.}\ \bibnamefont {Thompson}}, \
  and\ \bibinfo {author} {\bibfnamefont {T.}~\bibnamefont {Nishida}},\ }\href
  {\doibase 10.1063/1.2730561} {\bibfield  {journal} {\bibinfo  {journal}
  {Journal of Applied Physics}\ }\textbf {\bibinfo {volume} {101}},\ \bibinfo
  {pages} {104503} (\bibinfo {year} {2007})}\BibitemShut {NoStop}%
\bibitem [{\citenamefont {Sando}\ \emph {et~al.}(2014)\citenamefont {Sando},
  \citenamefont {Barth{\'{e}}l{\'{e}}my},\ and\ \citenamefont
  {Bibes}}]{Sando2014}%
  \BibitemOpen
  \bibfield  {author} {\bibinfo {author} {\bibfnamefont {D.}~\bibnamefont
  {Sando}}, \bibinfo {author} {\bibfnamefont {A.}~\bibnamefont
  {Barth{\'{e}}l{\'{e}}my}}, \ and\ \bibinfo {author} {\bibfnamefont
  {M.}~\bibnamefont {Bibes}},\ }\href {\doibase 10.1088/0953-8984/26/47/473201}
  {\bibfield  {journal} {\bibinfo  {journal} {Journal of Physics: Condensed
  Matter}\ }\textbf {\bibinfo {volume} {26}},\ \bibinfo {pages} {473201}
  (\bibinfo {year} {2014})}\BibitemShut {NoStop}%
\bibitem [{\citenamefont {Sando}\ \emph {et~al.}(2016)\citenamefont {Sando},
  \citenamefont {Xu}, \citenamefont {Bellaiche},\ and\ \citenamefont
  {Nagarajan}}]{Sando2016}%
  \BibitemOpen
  \bibfield  {author} {\bibinfo {author} {\bibfnamefont {D.}~\bibnamefont
  {Sando}}, \bibinfo {author} {\bibfnamefont {B.}~\bibnamefont {Xu}}, \bibinfo
  {author} {\bibfnamefont {L.}~\bibnamefont {Bellaiche}}, \ and\ \bibinfo
  {author} {\bibfnamefont {V.}~\bibnamefont {Nagarajan}},\ }\href {\doibase
  10.1063/1.4944558} {\bibfield  {journal} {\bibinfo  {journal} {Applied
  Physics Reviews}\ }\textbf {\bibinfo {volume} {3}},\ \bibinfo {pages}
  {011106} (\bibinfo {year} {2016})}\BibitemShut {NoStop}%
\bibitem [{\citenamefont {Yu}\ \emph {et~al.}(2012)\citenamefont {Yu},
  \citenamefont {Luo}, \citenamefont {Yi}, \citenamefont {Zhang}, \citenamefont
  {Rossell}, \citenamefont {Yang}, \citenamefont {You}, \citenamefont
  {Singh-Bhalla}, \citenamefont {Yang}, \citenamefont {He}, \citenamefont
  {Ramasse}, \citenamefont {Erni}, \citenamefont {Martin}, \citenamefont {Chu},
  \citenamefont {Pantelides}, \citenamefont {Pennycook},\ and\ \citenamefont
  {Ramesh}}]{Yu2012}%
  \BibitemOpen
  \bibfield  {author} {\bibinfo {author} {\bibfnamefont {P.}~\bibnamefont
  {Yu}}, \bibinfo {author} {\bibfnamefont {W.}~\bibnamefont {Luo}}, \bibinfo
  {author} {\bibfnamefont {D.}~\bibnamefont {Yi}}, \bibinfo {author}
  {\bibfnamefont {J.~X.}\ \bibnamefont {Zhang}}, \bibinfo {author}
  {\bibfnamefont {M.~D.}\ \bibnamefont {Rossell}}, \bibinfo {author}
  {\bibfnamefont {C.-H.}\ \bibnamefont {Yang}}, \bibinfo {author}
  {\bibfnamefont {L.}~\bibnamefont {You}}, \bibinfo {author} {\bibfnamefont
  {G.}~\bibnamefont {Singh-Bhalla}}, \bibinfo {author} {\bibfnamefont {S.~Y.}\
  \bibnamefont {Yang}}, \bibinfo {author} {\bibfnamefont {Q.}~\bibnamefont
  {He}}, \bibinfo {author} {\bibfnamefont {Q.~M.}\ \bibnamefont {Ramasse}},
  \bibinfo {author} {\bibfnamefont {R.}~\bibnamefont {Erni}}, \bibinfo {author}
  {\bibfnamefont {L.~W.}\ \bibnamefont {Martin}}, \bibinfo {author}
  {\bibfnamefont {Y.~H.}\ \bibnamefont {Chu}}, \bibinfo {author} {\bibfnamefont
  {S.~T.}\ \bibnamefont {Pantelides}}, \bibinfo {author} {\bibfnamefont
  {S.~J.}\ \bibnamefont {Pennycook}}, \ and\ \bibinfo {author} {\bibfnamefont
  {R.}~\bibnamefont {Ramesh}},\ }\href {\doibase 10.1073/pnas.1117990109}
  {\bibfield  {journal} {\bibinfo  {journal} {Proceedings of the National
  Academy of Sciences of the United States of America}\ }\textbf {\bibinfo
  {volume} {109}},\ \bibinfo {pages} {9710} (\bibinfo {year}
  {2012})}\BibitemShut {NoStop}%
\bibitem [{\citenamefont {Kim}\ \emph {et~al.}(2013)\citenamefont {Kim},
  \citenamefont {Kumar}, \citenamefont {Hatt}, \citenamefont {Morozovska},
  \citenamefont {Tselev}, \citenamefont {Biegalski}, \citenamefont {Ivanov},
  \citenamefont {Eliseev}, \citenamefont {Pennycook}, \citenamefont
  {Rondinelli}, \citenamefont {Kalinin},\ and\ \citenamefont
  {Borisevich}}]{Kim2013b}%
  \BibitemOpen
  \bibfield  {author} {\bibinfo {author} {\bibfnamefont {Y.-M.}\ \bibnamefont
  {Kim}}, \bibinfo {author} {\bibfnamefont {A.}~\bibnamefont {Kumar}}, \bibinfo
  {author} {\bibfnamefont {A.}~\bibnamefont {Hatt}}, \bibinfo {author}
  {\bibfnamefont {A.~N.}\ \bibnamefont {Morozovska}}, \bibinfo {author}
  {\bibfnamefont {A.}~\bibnamefont {Tselev}}, \bibinfo {author} {\bibfnamefont
  {M.~D.}\ \bibnamefont {Biegalski}}, \bibinfo {author} {\bibfnamefont
  {I.}~\bibnamefont {Ivanov}}, \bibinfo {author} {\bibfnamefont {E.~A.}\
  \bibnamefont {Eliseev}}, \bibinfo {author} {\bibfnamefont {S.~J.}\
  \bibnamefont {Pennycook}}, \bibinfo {author} {\bibfnamefont {J.~M.}\
  \bibnamefont {Rondinelli}}, \bibinfo {author} {\bibfnamefont {S.~V.}\
  \bibnamefont {Kalinin}}, \ and\ \bibinfo {author} {\bibfnamefont {A.~Y.}\
  \bibnamefont {Borisevich}},\ }\href {\doibase 10.1002/adma.201204584}
  {\bibfield  {journal} {\bibinfo  {journal} {Advanced Materials}\ }\textbf
  {\bibinfo {volume} {25}},\ \bibinfo {pages} {2497} (\bibinfo {year}
  {2013})}\BibitemShut {NoStop}%
\bibitem [{\citenamefont {Kresse}\ and\ \citenamefont
  {Hafner}(1993)}]{Kresse1993}%
  \BibitemOpen
  \bibfield  {author} {\bibinfo {author} {\bibfnamefont {G.}~\bibnamefont
  {Kresse}}\ and\ \bibinfo {author} {\bibfnamefont {J.}~\bibnamefont
  {Hafner}},\ }\href {\doibase 10.1103/PhysRevB.47.558} {\bibfield  {journal}
  {\bibinfo  {journal} {Physical Review B}\ }\textbf {\bibinfo {volume} {47}},\
  \bibinfo {pages} {558} (\bibinfo {year} {1993})}\BibitemShut {NoStop}%
\bibitem [{\citenamefont {Kresse}\ and\ \citenamefont
  {Hafner}(1994)}]{Kresse1994}%
  \BibitemOpen
  \bibfield  {author} {\bibinfo {author} {\bibfnamefont {G.}~\bibnamefont
  {Kresse}}\ and\ \bibinfo {author} {\bibfnamefont {J.}~\bibnamefont
  {Hafner}},\ }\href {\doibase 10.1103/PhysRevB.49.14251} {\bibfield  {journal}
  {\bibinfo  {journal} {Physical Review B}\ }\textbf {\bibinfo {volume} {49}},\
  \bibinfo {pages} {14251} (\bibinfo {year} {1994})}\BibitemShut {NoStop}%
\bibitem [{\citenamefont {Kresse}\ and\ \citenamefont
  {Furthm{\"{u}}ller}(1996{\natexlab{a}})}]{Kresse1996a}%
  \BibitemOpen
  \bibfield  {author} {\bibinfo {author} {\bibfnamefont {G.}~\bibnamefont
  {Kresse}}\ and\ \bibinfo {author} {\bibfnamefont {J.}~\bibnamefont
  {Furthm{\"{u}}ller}},\ }\href {\doibase 10.1016/0927-0256(96)00008-0}
  {\bibfield  {journal} {\bibinfo  {journal} {Computational Materials Science}\
  }\textbf {\bibinfo {volume} {6}},\ \bibinfo {pages} {15} (\bibinfo {year}
  {1996}{\natexlab{a}})}\BibitemShut {NoStop}%
\bibitem [{\citenamefont {Kresse}\ and\ \citenamefont
  {Furthm{\"{u}}ller}(1996{\natexlab{b}})}]{Kresse1996b}%
  \BibitemOpen
  \bibfield  {author} {\bibinfo {author} {\bibfnamefont {G.}~\bibnamefont
  {Kresse}}\ and\ \bibinfo {author} {\bibfnamefont {J.}~\bibnamefont
  {Furthm{\"{u}}ller}},\ }\href {\doibase 10.1103/PhysRevB.54.11169} {\bibfield
   {journal} {\bibinfo  {journal} {Physical Review B}\ }\textbf {\bibinfo
  {volume} {54}},\ \bibinfo {pages} {11169} (\bibinfo {year}
  {1996}{\natexlab{b}})}\BibitemShut {NoStop}%
\bibitem [{\citenamefont {Perdew}\ \emph {et~al.}(1996)\citenamefont {Perdew},
  \citenamefont {Burke},\ and\ \citenamefont {Ernzerhof}}]{Perdew1996}%
  \BibitemOpen
  \bibfield  {author} {\bibinfo {author} {\bibfnamefont {J.~P.}\ \bibnamefont
  {Perdew}}, \bibinfo {author} {\bibfnamefont {K.}~\bibnamefont {Burke}}, \
  and\ \bibinfo {author} {\bibfnamefont {M.}~\bibnamefont {Ernzerhof}},\ }\href
  {\doibase 10.1103/PhysRevLett.77.3865} {\bibfield  {journal} {\bibinfo
  {journal} {Physical Review Letters}\ }\textbf {\bibinfo {volume} {77}},\
  \bibinfo {pages} {3865} (\bibinfo {year} {1996})}\BibitemShut {NoStop}%
\bibitem [{\citenamefont {Dudarev}\ \emph {et~al.}(1998)\citenamefont
  {Dudarev}, \citenamefont {Savrasov}, \citenamefont {Humphreys}, \citenamefont
  {Sutton}, \citenamefont {Botton}, \citenamefont {Savrasov}, \citenamefont
  {Humphreys},\ and\ \citenamefont {Sutton}}]{Dudarev1998}%
  \BibitemOpen
  \bibfield  {author} {\bibinfo {author} {\bibfnamefont {S.~L.}\ \bibnamefont
  {Dudarev}}, \bibinfo {author} {\bibfnamefont {S.~Y.}\ \bibnamefont
  {Savrasov}}, \bibinfo {author} {\bibfnamefont {C.~J.}\ \bibnamefont
  {Humphreys}}, \bibinfo {author} {\bibfnamefont {A.~P.}\ \bibnamefont
  {Sutton}}, \bibinfo {author} {\bibfnamefont {G.~A.}\ \bibnamefont {Botton}},
  \bibinfo {author} {\bibfnamefont {S.~Y.}\ \bibnamefont {Savrasov}}, \bibinfo
  {author} {\bibfnamefont {C.~J.}\ \bibnamefont {Humphreys}}, \ and\ \bibinfo
  {author} {\bibfnamefont {A.~P.}\ \bibnamefont {Sutton}},\ }\href {\doibase
  10.1103/PhysRevB.57.1505} {\bibfield  {journal} {\bibinfo  {journal}
  {Physical Review B}\ }\textbf {\bibinfo {volume} {57}},\ \bibinfo {pages}
  {1505} (\bibinfo {year} {1998})}\BibitemShut {NoStop}%
\bibitem [{\citenamefont {Shenton}\ \emph {et~al.}(2017)\citenamefont
  {Shenton}, \citenamefont {Bowler},\ and\ \citenamefont
  {Cheah}}]{Shenton2017}%
  \BibitemOpen
  \bibfield  {author} {\bibinfo {author} {\bibfnamefont {J.~K.}\ \bibnamefont
  {Shenton}}, \bibinfo {author} {\bibfnamefont {D.~R.}\ \bibnamefont {Bowler}},
  \ and\ \bibinfo {author} {\bibfnamefont {W.~L.}\ \bibnamefont {Cheah}},\
  }\href {\doibase 10.1088/1361-648X/aa8935} {\bibfield  {journal} {\bibinfo
  {journal} {Journal of Physics: Condensed Matter}\ }\textbf {\bibinfo {volume}
  {29}},\ \bibinfo {pages} {445501} (\bibinfo {year} {2017})}\BibitemShut
  {NoStop}%
\bibitem [{\citenamefont {Bl{\"{o}}chl}(1994)}]{Blochl1994b}%
  \BibitemOpen
  \bibfield  {author} {\bibinfo {author} {\bibfnamefont {P.~E.}\ \bibnamefont
  {Bl{\"{o}}chl}},\ }\href {\doibase 10.1103/PhysRevB.50.17953} {\bibfield
  {journal} {\bibinfo  {journal} {Physical Review B}\ }\textbf {\bibinfo
  {volume} {50}},\ \bibinfo {pages} {17953} (\bibinfo {year}
  {1994})}\BibitemShut {NoStop}%
\bibitem [{\citenamefont {Kresse}\ and\ \citenamefont
  {Joubert}(1999)}]{Kresse1999b}%
  \BibitemOpen
  \bibfield  {author} {\bibinfo {author} {\bibfnamefont {G.}~\bibnamefont
  {Kresse}}\ and\ \bibinfo {author} {\bibfnamefont {D.}~\bibnamefont
  {Joubert}},\ }\href {\doibase 10.1103/PhysRevB.59.1758} {\bibfield  {journal}
  {\bibinfo  {journal} {Physical Review B}\ }\textbf {\bibinfo {volume} {59}},\
  \bibinfo {pages} {1758} (\bibinfo {year} {1999})}\BibitemShut {NoStop}%
\bibitem [{Note1()}]{Note1}%
  \BibitemOpen
  \bibinfo {note} {The Bi, Fe and O PAWs are dated: 6$\protect \mathrm {^{th}}$
  Sept. 2000, 2$\protect \mathrm {^{nd}}$ Aug. 2007 and 8$\protect \mathrm
  {^{th}}$ Apr. 2002 respectively}\BibitemShut {NoStop}%
\bibitem [{\citenamefont {Monkhorst}\ and\ \citenamefont
  {Pack}(1976)}]{Monkhorst1976}%
  \BibitemOpen
  \bibfield  {author} {\bibinfo {author} {\bibfnamefont {H.~J.}\ \bibnamefont
  {Monkhorst}}\ and\ \bibinfo {author} {\bibfnamefont {J.~D.}\ \bibnamefont
  {Pack}},\ }\href {\doibase 10.1103/PhysRevB.13.5188} {\bibfield  {journal}
  {\bibinfo  {journal} {Physical Review B}\ }\textbf {\bibinfo {volume} {13}},\
  \bibinfo {pages} {5188} (\bibinfo {year} {1976})}\BibitemShut {NoStop}%
\bibitem [{\citenamefont {Shenton}\ \emph {et~al.}(2018)\citenamefont
  {Shenton}, \citenamefont {Bowler},\ and\ \citenamefont
  {Cheah}}]{Shenton2018}%
  \BibitemOpen
  \bibfield  {author} {\bibinfo {author} {\bibfnamefont {J.~K.}\ \bibnamefont
  {Shenton}}, \bibinfo {author} {\bibfnamefont {D.~R.}\ \bibnamefont {Bowler}},
  \ and\ \bibinfo {author} {\bibfnamefont {W.~L.}\ \bibnamefont {Cheah}},\
  }\href {\doibase 10.5281/ZENODO.1198208} {\  (\bibinfo {year} {2018}),\
  10.5281/ZENODO.1198208}\BibitemShut {NoStop}%
\bibitem [{\citenamefont {King-Smith}\ and\ \citenamefont
  {Vanderbilt}(1993)}]{King-Smith1993}%
  \BibitemOpen
  \bibfield  {author} {\bibinfo {author} {\bibfnamefont {R.~D.}\ \bibnamefont
  {King-Smith}}\ and\ \bibinfo {author} {\bibfnamefont {D.}~\bibnamefont
  {Vanderbilt}},\ }\href {\doibase 10.1103/PhysRevB.47.1651} {\bibfield
  {journal} {\bibinfo  {journal} {Physical Review B}\ }\textbf {\bibinfo
  {volume} {47}},\ \bibinfo {pages} {1651} (\bibinfo {year}
  {1993})}\BibitemShut {NoStop}%
\bibitem [{\citenamefont {Resta}(1994)}]{Resta1994}%
  \BibitemOpen
  \bibfield  {author} {\bibinfo {author} {\bibfnamefont {R.}~\bibnamefont
  {Resta}},\ }\href {\doibase 10.1103/RevModPhys.66.899} {\bibfield  {journal}
  {\bibinfo  {journal} {Ferroelectrics}\ }\textbf {\bibinfo {volume} {151}},\
  \bibinfo {pages} {49} (\bibinfo {year} {1994})}\BibitemShut {NoStop}%
\bibitem [{\citenamefont {Neaton}\ \emph {et~al.}(2005)\citenamefont {Neaton},
  \citenamefont {Ederer}, \citenamefont {Waghmare}, \citenamefont {Spaldin},\
  and\ \citenamefont {Rabe}}]{Neaton2005}%
  \BibitemOpen
  \bibfield  {author} {\bibinfo {author} {\bibfnamefont {J.~B.}\ \bibnamefont
  {Neaton}}, \bibinfo {author} {\bibfnamefont {C.}~\bibnamefont {Ederer}},
  \bibinfo {author} {\bibfnamefont {U.~V.}\ \bibnamefont {Waghmare}}, \bibinfo
  {author} {\bibfnamefont {N.~A.}\ \bibnamefont {Spaldin}}, \ and\ \bibinfo
  {author} {\bibfnamefont {K.~M.}\ \bibnamefont {Rabe}},\ }\href {\doibase
  10.1103/PhysRevB.71.014113} {\bibfield  {journal} {\bibinfo  {journal}
  {Physical Review B}\ }\textbf {\bibinfo {volume} {71}},\ \bibinfo {pages}
  {014113} (\bibinfo {year} {2005})}\BibitemShut {NoStop}%
\bibitem [{Note2()}]{Note2}%
  \BibitemOpen
  \bibinfo {note} {\protect \leavevmode {\protect \color {blue}For a thorough
  discussion of alternative effective mass definitions, we refer the reader to
  the recent work of Whalley \protect \emph {et al.} \cite
  {Whalley2019}}}\BibitemShut {NoStop}%
\bibitem [{\citenamefont {Fonari}\ and\ \citenamefont
  {Sutton}(2012)}]{Fonari2012}%
  \BibitemOpen
  \bibfield  {author} {\bibinfo {author} {\bibfnamefont {A.}~\bibnamefont
  {Fonari}}\ and\ \bibinfo {author} {\bibfnamefont {C.}~\bibnamefont
  {Sutton}},\ }\href {https://github.com/afonari/emc} {\enquote {\bibinfo
  {title} {{Effective Mass Calculator (v1.50)}},}\ } (\bibinfo {year}
  {2012})\BibitemShut {NoStop}%
\bibitem [{\citenamefont {Sosnowska}\ \emph {et~al.}(1982)\citenamefont
  {Sosnowska}, \citenamefont {Neumaier},\ and\ \citenamefont
  {Steichele}}]{Sosnowska1982b}%
  \BibitemOpen
  \bibfield  {author} {\bibinfo {author} {\bibfnamefont {I.}~\bibnamefont
  {Sosnowska}}, \bibinfo {author} {\bibfnamefont {T.~P.}\ \bibnamefont
  {Neumaier}}, \ and\ \bibinfo {author} {\bibfnamefont {E.}~\bibnamefont
  {Steichele}},\ }\href {\doibase 10.1088/0022-3719/15/23/020} {\bibfield
  {journal} {\bibinfo  {journal} {Journal of Physics C: Solid State Physics}\
  }\textbf {\bibinfo {volume} {15}},\ \bibinfo {pages} {4835} (\bibinfo {year}
  {1982})}\BibitemShut {NoStop}%
\bibitem [{\citenamefont {Di{\'{e}}guez}\ \emph {et~al.}(2011)\citenamefont
  {Di{\'{e}}guez}, \citenamefont {Gonz{\'{a}}lez-V{\'{a}}zquez}, \citenamefont
  {Wojde{\l}},\ and\ \citenamefont {{\'{I}}{\~{n}}iguez}}]{Dieguez2011}%
  \BibitemOpen
  \bibfield  {author} {\bibinfo {author} {\bibfnamefont {O.}~\bibnamefont
  {Di{\'{e}}guez}}, \bibinfo {author} {\bibfnamefont {O.~E.}\ \bibnamefont
  {Gonz{\'{a}}lez-V{\'{a}}zquez}}, \bibinfo {author} {\bibfnamefont {J.~C.}\
  \bibnamefont {Wojde{\l}}}, \ and\ \bibinfo {author} {\bibfnamefont
  {J.}~\bibnamefont {{\'{I}}{\~{n}}iguez}},\ }\href {\doibase
  10.1103/PhysRevB.83.094105} {\bibfield  {journal} {\bibinfo  {journal}
  {Physical Review B}\ }\textbf {\bibinfo {volume} {83}},\ \bibinfo {pages}
  {94105} (\bibinfo {year} {2011})}\BibitemShut {NoStop}%
\bibitem [{\citenamefont {Hatt}\ \emph {et~al.}(2010)\citenamefont {Hatt},
  \citenamefont {Spaldin},\ and\ \citenamefont {Ederer}}]{Hatt2010}%
  \BibitemOpen
  \bibfield  {author} {\bibinfo {author} {\bibfnamefont {A.~J.}\ \bibnamefont
  {Hatt}}, \bibinfo {author} {\bibfnamefont {N.~A.}\ \bibnamefont {Spaldin}}, \
  and\ \bibinfo {author} {\bibfnamefont {C.}~\bibnamefont {Ederer}},\ }\href
  {\doibase 10.1103/PhysRevB.81.054109} {\bibfield  {journal} {\bibinfo
  {journal} {Physical Review B}\ }\textbf {\bibinfo {volume} {81}},\ \bibinfo
  {pages} {054109} (\bibinfo {year} {2010})}\BibitemShut {NoStop}%
\bibitem [{\citenamefont {Ederer}\ and\ \citenamefont
  {Spaldin}(2005{\natexlab{a}})}]{Ederer2005b}%
  \BibitemOpen
  \bibfield  {author} {\bibinfo {author} {\bibfnamefont {C.}~\bibnamefont
  {Ederer}}\ and\ \bibinfo {author} {\bibfnamefont {N.~A.}\ \bibnamefont
  {Spaldin}},\ }\href {\doibase 10.1103/PhysRevB.71.224103} {\bibfield
  {journal} {\bibinfo  {journal} {Physical Review B}\ }\textbf {\bibinfo
  {volume} {71}},\ \bibinfo {pages} {224103} (\bibinfo {year}
  {2005}{\natexlab{a}})}\BibitemShut {NoStop}%
\bibitem [{\citenamefont {Ravindran}\ \emph {et~al.}(2006)\citenamefont
  {Ravindran}, \citenamefont {Vidya}, \citenamefont {Kjekshus}, \citenamefont
  {Fjellv{\aa}g},\ and\ \citenamefont {Eriksson}}]{Ravindran2006}%
  \BibitemOpen
  \bibfield  {author} {\bibinfo {author} {\bibfnamefont {P.}~\bibnamefont
  {Ravindran}}, \bibinfo {author} {\bibfnamefont {R.}~\bibnamefont {Vidya}},
  \bibinfo {author} {\bibfnamefont {A.}~\bibnamefont {Kjekshus}}, \bibinfo
  {author} {\bibfnamefont {H.}~\bibnamefont {Fjellv{\aa}g}}, \ and\ \bibinfo
  {author} {\bibfnamefont {O.}~\bibnamefont {Eriksson}},\ }\href {\doibase
  10.1103/PhysRevB.74.224412} {\bibfield  {journal} {\bibinfo  {journal}
  {Physical Review B}\ }\textbf {\bibinfo {volume} {74}},\ \bibinfo {pages}
  {224412} (\bibinfo {year} {2006})}\BibitemShut {NoStop}%
\bibitem [{\citenamefont {Ederer}\ and\ \citenamefont
  {Spaldin}(2005{\natexlab{b}})}]{Ederer2005a}%
  \BibitemOpen
  \bibfield  {author} {\bibinfo {author} {\bibfnamefont {C.}~\bibnamefont
  {Ederer}}\ and\ \bibinfo {author} {\bibfnamefont {N.~A.}\ \bibnamefont
  {Spaldin}},\ }\href {\doibase 10.1103/PhysRevLett.95.257601} {\bibfield
  {journal} {\bibinfo  {journal} {Physical Review Letters}\ }\textbf {\bibinfo
  {volume} {95}},\ \bibinfo {pages} {257601} (\bibinfo {year}
  {2005}{\natexlab{b}})}\BibitemShut {NoStop}%
\bibitem [{\citenamefont {Zhang}\ \emph {et~al.}(2011)\citenamefont {Zhang},
  \citenamefont {He}, \citenamefont {Trassin}, \citenamefont {Luo},
  \citenamefont {Yi}, \citenamefont {Rossell}, \citenamefont {Yu},
  \citenamefont {You}, \citenamefont {Wang}, \citenamefont {Kuo}, \citenamefont
  {Heron}, \citenamefont {Hu}, \citenamefont {Zeches}, \citenamefont {Lin},
  \citenamefont {Tanaka}, \citenamefont {Chen}, \citenamefont {Tjeng},
  \citenamefont {Chu},\ and\ \citenamefont {Ramesh}}]{Zhang2011}%
  \BibitemOpen
  \bibfield  {author} {\bibinfo {author} {\bibfnamefont {J.~X.}\ \bibnamefont
  {Zhang}}, \bibinfo {author} {\bibfnamefont {Q.}~\bibnamefont {He}}, \bibinfo
  {author} {\bibfnamefont {M.}~\bibnamefont {Trassin}}, \bibinfo {author}
  {\bibfnamefont {W.}~\bibnamefont {Luo}}, \bibinfo {author} {\bibfnamefont
  {D.}~\bibnamefont {Yi}}, \bibinfo {author} {\bibfnamefont {M.~D.}\
  \bibnamefont {Rossell}}, \bibinfo {author} {\bibfnamefont {P.}~\bibnamefont
  {Yu}}, \bibinfo {author} {\bibfnamefont {L.}~\bibnamefont {You}}, \bibinfo
  {author} {\bibfnamefont {C.~H.}\ \bibnamefont {Wang}}, \bibinfo {author}
  {\bibfnamefont {C.~Y.}\ \bibnamefont {Kuo}}, \bibinfo {author} {\bibfnamefont
  {J.~T.}\ \bibnamefont {Heron}}, \bibinfo {author} {\bibfnamefont
  {Z.}~\bibnamefont {Hu}}, \bibinfo {author} {\bibfnamefont {R.~J.}\
  \bibnamefont {Zeches}}, \bibinfo {author} {\bibfnamefont {H.~J.}\
  \bibnamefont {Lin}}, \bibinfo {author} {\bibfnamefont {A.}~\bibnamefont
  {Tanaka}}, \bibinfo {author} {\bibfnamefont {C.~T.}\ \bibnamefont {Chen}},
  \bibinfo {author} {\bibfnamefont {L.~H.}\ \bibnamefont {Tjeng}}, \bibinfo
  {author} {\bibfnamefont {Y.~H.}\ \bibnamefont {Chu}}, \ and\ \bibinfo
  {author} {\bibfnamefont {R.}~\bibnamefont {Ramesh}},\ }\href {\doibase
  10.1103/PhysRevLett.107.147602} {\bibfield  {journal} {\bibinfo  {journal}
  {Physical Review Letters}\ }\textbf {\bibinfo {volume} {107}},\ \bibinfo
  {pages} {147602} (\bibinfo {year} {2011})}\BibitemShut {NoStop}%
\bibitem [{\citenamefont {Glazer}(1972)}]{Glazer1972a}%
  \BibitemOpen
  \bibfield  {author} {\bibinfo {author} {\bibfnamefont {A.~M.}\ \bibnamefont
  {Glazer}},\ }\href {\doibase 10.1107/S0567740872007976} {\bibfield  {journal}
  {\bibinfo  {journal} {Acta Crystallographica Section B: Structural
  Crystallography and Crystal Chemistry}\ }\textbf {\bibinfo {volume} {28}},\
  \bibinfo {pages} {3384} (\bibinfo {year} {1972})}\BibitemShut {NoStop}%
\bibitem [{\citenamefont {Rana}\ \emph {et~al.}(2007)\citenamefont {Rana},
  \citenamefont {Takahashi}, \citenamefont {Mavani}, \citenamefont {Kawayama},
  \citenamefont {Murakami}, \citenamefont {Tonouchi}, \citenamefont {Yanagida},
  \citenamefont {Tanaka},\ and\ \citenamefont {Kawai}}]{Rana2007}%
  \BibitemOpen
  \bibfield  {author} {\bibinfo {author} {\bibfnamefont {D.~S.}\ \bibnamefont
  {Rana}}, \bibinfo {author} {\bibfnamefont {K.}~\bibnamefont {Takahashi}},
  \bibinfo {author} {\bibfnamefont {K.~R.}\ \bibnamefont {Mavani}}, \bibinfo
  {author} {\bibfnamefont {I.}~\bibnamefont {Kawayama}}, \bibinfo {author}
  {\bibfnamefont {H.}~\bibnamefont {Murakami}}, \bibinfo {author}
  {\bibfnamefont {M.}~\bibnamefont {Tonouchi}}, \bibinfo {author}
  {\bibfnamefont {T.}~\bibnamefont {Yanagida}}, \bibinfo {author}
  {\bibfnamefont {H.}~\bibnamefont {Tanaka}}, \ and\ \bibinfo {author}
  {\bibfnamefont {T.}~\bibnamefont {Kawai}},\ }\href {\doibase
  10.1103/PhysRevB.75.060405} {\bibfield  {journal} {\bibinfo  {journal}
  {Physical Review B}\ }\textbf {\bibinfo {volume} {75}},\ \bibinfo {pages}
  {060405} (\bibinfo {year} {2007})}\BibitemShut {NoStop}%
\bibitem [{\citenamefont {Infante}\ \emph {et~al.}(2010)\citenamefont
  {Infante}, \citenamefont {Lisenkov}, \citenamefont {Dup{\'{e}}},
  \citenamefont {Bibes}, \citenamefont {Fusil}, \citenamefont {Jacquet},
  \citenamefont {Geneste}, \citenamefont {Petit}, \citenamefont {Courtial},
  \citenamefont {Juraszek}, \citenamefont {Bellaiche}, \citenamefont
  {Barth{\'{e}}l{\'{e}}my},\ and\ \citenamefont {Dkhil}}]{Infante2010}%
  \BibitemOpen
  \bibfield  {author} {\bibinfo {author} {\bibfnamefont {I.~C.}\ \bibnamefont
  {Infante}}, \bibinfo {author} {\bibfnamefont {S.}~\bibnamefont {Lisenkov}},
  \bibinfo {author} {\bibfnamefont {B.}~\bibnamefont {Dup{\'{e}}}}, \bibinfo
  {author} {\bibfnamefont {M.}~\bibnamefont {Bibes}}, \bibinfo {author}
  {\bibfnamefont {S.}~\bibnamefont {Fusil}}, \bibinfo {author} {\bibfnamefont
  {E.}~\bibnamefont {Jacquet}}, \bibinfo {author} {\bibfnamefont
  {G.}~\bibnamefont {Geneste}}, \bibinfo {author} {\bibfnamefont
  {S.}~\bibnamefont {Petit}}, \bibinfo {author} {\bibfnamefont
  {A.}~\bibnamefont {Courtial}}, \bibinfo {author} {\bibfnamefont
  {J.}~\bibnamefont {Juraszek}}, \bibinfo {author} {\bibfnamefont
  {L.}~\bibnamefont {Bellaiche}}, \bibinfo {author} {\bibfnamefont
  {A.}~\bibnamefont {Barth{\'{e}}l{\'{e}}my}}, \ and\ \bibinfo {author}
  {\bibfnamefont {B.}~\bibnamefont {Dkhil}},\ }\href {\doibase
  10.1103/PhysRevLett.105.057601} {\bibfield  {journal} {\bibinfo  {journal}
  {Physical Review Letters}\ }\textbf {\bibinfo {volume} {105}},\ \bibinfo
  {pages} {057601} (\bibinfo {year} {2010})}\BibitemShut {NoStop}%
\bibitem [{\citenamefont {Sando}\ \emph {et~al.}(2013)\citenamefont {Sando},
  \citenamefont {Agbelele}, \citenamefont {Rahmedov}, \citenamefont {Liu},
  \citenamefont {Rovillain}, \citenamefont {Toulouse}, \citenamefont {Infante},
  \citenamefont {Pyatakov}, \citenamefont {Fusil}, \citenamefont {Jacquet},
  \citenamefont {Carr{\'{e}}t{\'{e}}ro}, \citenamefont {Deranlot},
  \citenamefont {Lisenkov}, \citenamefont {Wang}, \citenamefont {{Le Breton}},
  \citenamefont {Cazayous}, \citenamefont {Sacuto}, \citenamefont {Juraszek},
  \citenamefont {Zvezdin}, \citenamefont {Bellaiche}, \citenamefont {Dkhil},
  \citenamefont {Barth{\'{e}}l{\'{e}}my},\ and\ \citenamefont
  {Bibes}}]{Sando2013}%
  \BibitemOpen
  \bibfield  {author} {\bibinfo {author} {\bibfnamefont {D.}~\bibnamefont
  {Sando}}, \bibinfo {author} {\bibfnamefont {A.}~\bibnamefont {Agbelele}},
  \bibinfo {author} {\bibfnamefont {D.}~\bibnamefont {Rahmedov}}, \bibinfo
  {author} {\bibfnamefont {J.}~\bibnamefont {Liu}}, \bibinfo {author}
  {\bibfnamefont {P.}~\bibnamefont {Rovillain}}, \bibinfo {author}
  {\bibfnamefont {C.}~\bibnamefont {Toulouse}}, \bibinfo {author}
  {\bibfnamefont {I.~C.}\ \bibnamefont {Infante}}, \bibinfo {author}
  {\bibfnamefont {A.~P.}\ \bibnamefont {Pyatakov}}, \bibinfo {author}
  {\bibfnamefont {S.}~\bibnamefont {Fusil}}, \bibinfo {author} {\bibfnamefont
  {E.}~\bibnamefont {Jacquet}}, \bibinfo {author} {\bibfnamefont
  {C.}~\bibnamefont {Carr{\'{e}}t{\'{e}}ro}}, \bibinfo {author} {\bibfnamefont
  {C.}~\bibnamefont {Deranlot}}, \bibinfo {author} {\bibfnamefont
  {S.}~\bibnamefont {Lisenkov}}, \bibinfo {author} {\bibfnamefont
  {D.}~\bibnamefont {Wang}}, \bibinfo {author} {\bibfnamefont {J.-M.}\
  \bibnamefont {{Le Breton}}}, \bibinfo {author} {\bibfnamefont
  {M.}~\bibnamefont {Cazayous}}, \bibinfo {author} {\bibfnamefont
  {A.}~\bibnamefont {Sacuto}}, \bibinfo {author} {\bibfnamefont
  {J.}~\bibnamefont {Juraszek}}, \bibinfo {author} {\bibfnamefont {A.~K.}\
  \bibnamefont {Zvezdin}}, \bibinfo {author} {\bibfnamefont {L.}~\bibnamefont
  {Bellaiche}}, \bibinfo {author} {\bibfnamefont {B.}~\bibnamefont {Dkhil}},
  \bibinfo {author} {\bibfnamefont {A.}~\bibnamefont {Barth{\'{e}}l{\'{e}}my}},
  \ and\ \bibinfo {author} {\bibfnamefont {M.}~\bibnamefont {Bibes}},\ }\href
  {\doibase 10.1038/nmat3629} {\bibfield  {journal} {\bibinfo  {journal}
  {Nature Materials}\ }\textbf {\bibinfo {volume} {12}},\ \bibinfo {pages}
  {641} (\bibinfo {year} {2013})}\BibitemShut {NoStop}%
\bibitem [{Note3()}]{Note3}%
  \BibitemOpen
  \bibinfo {note} {Because we project the wavefunctions onto $d$ orbitals
  defined with respect to the \protect \emph {global} Cartesian axes, the
  ability to resolve the difference between $t_{2g}$ and $e_g$ states
  diminishes as the local octahedral axes rotate relative to the global
  Cartesian axes. In other words, the `mixing' of the $t_{2g}$ and $e_{g}$
  manifolds may be an artefact of the way the projections are done. To
  unambiguously resolve the relative contributions of Fe $t_{2g}$ and $e_{g}$
  states, one might perform a transformation from the global Cartesian basis to
  one that is local to each octahedron, as was done in Ref. \protect \citenum
  {Mellan2015}. However, knowledge of the precise level of mixing between the
  Fe $t_{2g}$ and $e_{g}$ states goes beyond the requirements for the present
  study.}\BibitemShut {Stop}%
\bibitem [{\citenamefont {Yang}\ \emph {et~al.}(2009)\citenamefont {Yang},
  \citenamefont {Martin}, \citenamefont {Byrnes}, \citenamefont {Conry},
  \citenamefont {Basu}, \citenamefont {Paran}, \citenamefont {Reichertz},
  \citenamefont {Ihlefeld}, \citenamefont {Adamo}, \citenamefont {Melville},
  \citenamefont {Chu}, \citenamefont {Yang}, \citenamefont {Musfeldt},
  \citenamefont {Schlom}, \citenamefont {Ager},\ and\ \citenamefont
  {Ramesh}}]{Yang2009b}%
  \BibitemOpen
  \bibfield  {author} {\bibinfo {author} {\bibfnamefont {S.~Y.}\ \bibnamefont
  {Yang}}, \bibinfo {author} {\bibfnamefont {L.~W.}\ \bibnamefont {Martin}},
  \bibinfo {author} {\bibfnamefont {S.~J.}\ \bibnamefont {Byrnes}}, \bibinfo
  {author} {\bibfnamefont {T.~E.}\ \bibnamefont {Conry}}, \bibinfo {author}
  {\bibfnamefont {S.~R.}\ \bibnamefont {Basu}}, \bibinfo {author}
  {\bibfnamefont {D.}~\bibnamefont {Paran}}, \bibinfo {author} {\bibfnamefont
  {L.}~\bibnamefont {Reichertz}}, \bibinfo {author} {\bibfnamefont
  {J.}~\bibnamefont {Ihlefeld}}, \bibinfo {author} {\bibfnamefont
  {C.}~\bibnamefont {Adamo}}, \bibinfo {author} {\bibfnamefont
  {A.}~\bibnamefont {Melville}}, \bibinfo {author} {\bibfnamefont {Y.-H.}\
  \bibnamefont {Chu}}, \bibinfo {author} {\bibfnamefont {C.-H.}\ \bibnamefont
  {Yang}}, \bibinfo {author} {\bibfnamefont {J.~L.}\ \bibnamefont {Musfeldt}},
  \bibinfo {author} {\bibfnamefont {D.~G.}\ \bibnamefont {Schlom}}, \bibinfo
  {author} {\bibfnamefont {J.~W.}\ \bibnamefont {Ager}}, \ and\ \bibinfo
  {author} {\bibfnamefont {R.}~\bibnamefont {Ramesh}},\ }\href {\doibase
  10.1063/1.3204695} {\bibfield  {journal} {\bibinfo  {journal} {Applied
  Physics Letters}\ }\textbf {\bibinfo {volume} {95}},\ \bibinfo {pages}
  {062909} (\bibinfo {year} {2009})}\BibitemShut {NoStop}%
\bibitem [{\citenamefont {Ji}\ \emph {et~al.}(2010)\citenamefont {Ji},
  \citenamefont {Yao},\ and\ \citenamefont {Liang}}]{Ji2010}%
  \BibitemOpen
  \bibfield  {author} {\bibinfo {author} {\bibfnamefont {W.}~\bibnamefont
  {Ji}}, \bibinfo {author} {\bibfnamefont {K.}~\bibnamefont {Yao}}, \ and\
  \bibinfo {author} {\bibfnamefont {Y.~C.}\ \bibnamefont {Liang}},\ }\href
  {\doibase 10.1002/adma.200902985} {\bibfield  {journal} {\bibinfo  {journal}
  {Advanced Materials}\ }\textbf {\bibinfo {volume} {22}},\ \bibinfo {pages}
  {1763} (\bibinfo {year} {2010})}\BibitemShut {NoStop}%
\bibitem [{\citenamefont {Paillard}\ \emph {et~al.}(2016)\citenamefont
  {Paillard}, \citenamefont {Bai}, \citenamefont {Infante}, \citenamefont
  {Guennou}, \citenamefont {Geneste}, \citenamefont {Alexe}, \citenamefont
  {Kreisel},\ and\ \citenamefont {Dkhil}}]{Paillard2016}%
  \BibitemOpen
  \bibfield  {author} {\bibinfo {author} {\bibfnamefont {C.}~\bibnamefont
  {Paillard}}, \bibinfo {author} {\bibfnamefont {X.}~\bibnamefont {Bai}},
  \bibinfo {author} {\bibfnamefont {I.~C.}\ \bibnamefont {Infante}}, \bibinfo
  {author} {\bibfnamefont {M.}~\bibnamefont {Guennou}}, \bibinfo {author}
  {\bibfnamefont {G.}~\bibnamefont {Geneste}}, \bibinfo {author} {\bibfnamefont
  {M.}~\bibnamefont {Alexe}}, \bibinfo {author} {\bibfnamefont
  {J.}~\bibnamefont {Kreisel}}, \ and\ \bibinfo {author} {\bibfnamefont
  {B.}~\bibnamefont {Dkhil}},\ }\href {\doibase 10.1002/adma.201505215}
  {\bibfield  {journal} {\bibinfo  {journal} {Advanced Materials}\ }\textbf
  {\bibinfo {volume} {28}},\ \bibinfo {pages} {5153} (\bibinfo {year}
  {2016})}\BibitemShut {NoStop}%
\bibitem [{\citenamefont {Wu}\ \emph {et~al.}(2014)\citenamefont {Wu},
  \citenamefont {Guo}, \citenamefont {Zhang}, \citenamefont {Duan},
  \citenamefont {Li},\ and\ \citenamefont {Liu}}]{Wu2014}%
  \BibitemOpen
  \bibfield  {author} {\bibinfo {author} {\bibfnamefont {F.}~\bibnamefont
  {Wu}}, \bibinfo {author} {\bibfnamefont {Y.}~\bibnamefont {Guo}}, \bibinfo
  {author} {\bibfnamefont {Y.}~\bibnamefont {Zhang}}, \bibinfo {author}
  {\bibfnamefont {H.}~\bibnamefont {Duan}}, \bibinfo {author} {\bibfnamefont
  {H.}~\bibnamefont {Li}}, \ and\ \bibinfo {author} {\bibfnamefont
  {H.}~\bibnamefont {Liu}},\ }\href {\doibase 10.1021/jp5059462} {\bibfield
  {journal} {\bibinfo  {journal} {The Journal of Physical Chemistry C}\
  }\textbf {\bibinfo {volume} {118}},\ \bibinfo {pages} {15200} (\bibinfo
  {year} {2014})}\BibitemShut {NoStop}%
\bibitem [{\citenamefont {Borisevich}\ \emph {et~al.}(2010)\citenamefont
  {Borisevich}, \citenamefont {Chang}, \citenamefont {Huijben}, \citenamefont
  {Oxley}, \citenamefont {Okamoto}, \citenamefont {Niranjan}, \citenamefont
  {Burton}, \citenamefont {Tsymbal}, \citenamefont {Chu}, \citenamefont {Yu},
  \citenamefont {Ramesh}, \citenamefont {Kalinin},\ and\ \citenamefont
  {Pennycook}}]{Borisevich2010}%
  \BibitemOpen
  \bibfield  {author} {\bibinfo {author} {\bibfnamefont {A.~Y.}\ \bibnamefont
  {Borisevich}}, \bibinfo {author} {\bibfnamefont {H.~J.}\ \bibnamefont
  {Chang}}, \bibinfo {author} {\bibfnamefont {M.}~\bibnamefont {Huijben}},
  \bibinfo {author} {\bibfnamefont {M.~P.}\ \bibnamefont {Oxley}}, \bibinfo
  {author} {\bibfnamefont {S.}~\bibnamefont {Okamoto}}, \bibinfo {author}
  {\bibfnamefont {M.~K.}\ \bibnamefont {Niranjan}}, \bibinfo {author}
  {\bibfnamefont {J.~D.}\ \bibnamefont {Burton}}, \bibinfo {author}
  {\bibfnamefont {E.~Y.}\ \bibnamefont {Tsymbal}}, \bibinfo {author}
  {\bibfnamefont {Y.~H.}\ \bibnamefont {Chu}}, \bibinfo {author} {\bibfnamefont
  {P.}~\bibnamefont {Yu}}, \bibinfo {author} {\bibfnamefont {R.}~\bibnamefont
  {Ramesh}}, \bibinfo {author} {\bibfnamefont {S.~V.}\ \bibnamefont {Kalinin}},
  \ and\ \bibinfo {author} {\bibfnamefont {S.~J.}\ \bibnamefont {Pennycook}},\
  }\href {\doibase 10.1103/PhysRevLett.105.087204} {\bibfield  {journal}
  {\bibinfo  {journal} {Physical Review Letters}\ }\textbf {\bibinfo {volume}
  {105}},\ \bibinfo {pages} {087204} (\bibinfo {year} {2010})}\BibitemShut
  {NoStop}%
\bibitem [{\citenamefont {He}\ \emph {et~al.}(2010)\citenamefont {He},
  \citenamefont {Borisevich}, \citenamefont {Kalinin}, \citenamefont
  {Pennycook},\ and\ \citenamefont {Pantelides}}]{He2010}%
  \BibitemOpen
  \bibfield  {author} {\bibinfo {author} {\bibfnamefont {J.}~\bibnamefont
  {He}}, \bibinfo {author} {\bibfnamefont {A.}~\bibnamefont {Borisevich}},
  \bibinfo {author} {\bibfnamefont {S.~V.}\ \bibnamefont {Kalinin}}, \bibinfo
  {author} {\bibfnamefont {S.~J.}\ \bibnamefont {Pennycook}}, \ and\ \bibinfo
  {author} {\bibfnamefont {S.~T.}\ \bibnamefont {Pantelides}},\ }\href
  {\doibase 10.1103/PhysRevLett.105.227203} {\bibfield  {journal} {\bibinfo
  {journal} {Physical Review Letters}\ }\textbf {\bibinfo {volume} {105}},\
  \bibinfo {pages} {227203} (\bibinfo {year} {2010})}\BibitemShut {NoStop}%
\bibitem [{\citenamefont {Kan}\ \emph {et~al.}(2016)\citenamefont {Kan},
  \citenamefont {Aso}, \citenamefont {Sato}, \citenamefont {Haruta},
  \citenamefont {Kurata},\ and\ \citenamefont {Shimakawa}}]{Kan2016}%
  \BibitemOpen
  \bibfield  {author} {\bibinfo {author} {\bibfnamefont {D.}~\bibnamefont
  {Kan}}, \bibinfo {author} {\bibfnamefont {R.}~\bibnamefont {Aso}}, \bibinfo
  {author} {\bibfnamefont {R.}~\bibnamefont {Sato}}, \bibinfo {author}
  {\bibfnamefont {M.}~\bibnamefont {Haruta}}, \bibinfo {author} {\bibfnamefont
  {H.}~\bibnamefont {Kurata}}, \ and\ \bibinfo {author} {\bibfnamefont
  {Y.}~\bibnamefont {Shimakawa}},\ }\href {\doibase 10.1038/nmat4580}
  {\bibfield  {journal} {\bibinfo  {journal} {Nature Materials}\ }\textbf
  {\bibinfo {volume} {15}},\ \bibinfo {pages} {432} (\bibinfo {year}
  {2016})}\BibitemShut {NoStop}%
\bibitem [{\citenamefont {Rondinelli}\ \emph {et~al.}(2012)\citenamefont
  {Rondinelli}, \citenamefont {May},\ and\ \citenamefont
  {Freeland}}]{Rondinelli2012}%
  \BibitemOpen
  \bibfield  {author} {\bibinfo {author} {\bibfnamefont {J.~M.}\ \bibnamefont
  {Rondinelli}}, \bibinfo {author} {\bibfnamefont {S.~J.}\ \bibnamefont {May}},
  \ and\ \bibinfo {author} {\bibfnamefont {J.~W.}\ \bibnamefont {Freeland}},\
  }\href {\doibase 10.1557/mrs.2012.49} {\bibfield  {journal} {\bibinfo
  {journal} {MRS Bulletin}\ }\textbf {\bibinfo {volume} {37}},\ \bibinfo
  {pages} {261} (\bibinfo {year} {2012})}\BibitemShut {NoStop}%
\bibitem [{\citenamefont {Chen}\ \emph {et~al.}(2010)\citenamefont {Chen},
  \citenamefont {Podraza}, \citenamefont {Xu}, \citenamefont {Melville},
  \citenamefont {Vlahos}, \citenamefont {Gopalan}, \citenamefont {Ramesh},
  \citenamefont {Schlom},\ and\ \citenamefont {Musfeldt}}]{Chen2010}%
  \BibitemOpen
  \bibfield  {author} {\bibinfo {author} {\bibfnamefont {P.}~\bibnamefont
  {Chen}}, \bibinfo {author} {\bibfnamefont {N.~J.}\ \bibnamefont {Podraza}},
  \bibinfo {author} {\bibfnamefont {X.~S.}\ \bibnamefont {Xu}}, \bibinfo
  {author} {\bibfnamefont {A.}~\bibnamefont {Melville}}, \bibinfo {author}
  {\bibfnamefont {E.}~\bibnamefont {Vlahos}}, \bibinfo {author} {\bibfnamefont
  {V.}~\bibnamefont {Gopalan}}, \bibinfo {author} {\bibfnamefont
  {R.}~\bibnamefont {Ramesh}}, \bibinfo {author} {\bibfnamefont {D.~G.}\
  \bibnamefont {Schlom}}, \ and\ \bibinfo {author} {\bibfnamefont {J.~L.}\
  \bibnamefont {Musfeldt}},\ }\href {\doibase 10.1063/1.3364133} {\bibfield
  {journal} {\bibinfo  {journal} {Applied Physics Letters}\ }\textbf {\bibinfo
  {volume} {96}},\ \bibinfo {pages} {131907} (\bibinfo {year}
  {2010})}\BibitemShut {NoStop}%
\bibitem [{\citenamefont {Wang}\ \emph {et~al.}(2015)\citenamefont {Wang},
  \citenamefont {Tan},\ and\ \citenamefont {Liu}}]{Wang2015w}%
  \BibitemOpen
  \bibfield  {author} {\bibinfo {author} {\bibfnamefont {Q.}~\bibnamefont
  {Wang}}, \bibinfo {author} {\bibfnamefont {Q.}~\bibnamefont {Tan}}, \ and\
  \bibinfo {author} {\bibfnamefont {Y.}~\bibnamefont {Liu}},\ }\href {\doibase
  10.1016/j.commatsci.2015.04.015} {\bibfield  {journal} {\bibinfo  {journal}
  {Computational Materials Science}\ }\textbf {\bibinfo {volume} {105}},\
  \bibinfo {pages} {1} (\bibinfo {year} {2015})}\BibitemShut {NoStop}%
\bibitem [{\citenamefont {Rong}\ \emph {et~al.}(2015)\citenamefont {Rong},
  \citenamefont {Wang}, \citenamefont {Xiao},\ and\ \citenamefont
  {Xu}}]{Rong2015}%
  \BibitemOpen
  \bibfield  {author} {\bibinfo {author} {\bibfnamefont {Q.-Y.}\ \bibnamefont
  {Rong}}, \bibinfo {author} {\bibfnamefont {L.-L.}\ \bibnamefont {Wang}},
  \bibinfo {author} {\bibfnamefont {W.-Z.}\ \bibnamefont {Xiao}}, \ and\
  \bibinfo {author} {\bibfnamefont {L.}~\bibnamefont {Xu}},\ }\href {\doibase
  10.1016/j.physb.2014.08.028} {\bibfield  {journal} {\bibinfo  {journal}
  {Physica B: Condensed Matter}\ }\textbf {\bibinfo {volume} {457}},\ \bibinfo
  {pages} {1} (\bibinfo {year} {2015})}\BibitemShut {NoStop}%
\bibitem [{\citenamefont {Bardeen}\ and\ \citenamefont
  {Shockley}(1950)}]{Bardeen1950}%
  \BibitemOpen
  \bibfield  {author} {\bibinfo {author} {\bibfnamefont {J.}~\bibnamefont
  {Bardeen}}\ and\ \bibinfo {author} {\bibfnamefont {W.}~\bibnamefont
  {Shockley}},\ }\href {\doibase 10.1103/PhysRev.80.72} {\bibfield  {journal}
  {\bibinfo  {journal} {Physical Review}\ }\textbf {\bibinfo {volume} {80}},\
  \bibinfo {pages} {72} (\bibinfo {year} {1950})}\BibitemShut {NoStop}%
\bibitem [{\citenamefont {Madsen}\ and\ \citenamefont
  {Singh}(2006)}]{Madsen2006}%
  \BibitemOpen
  \bibfield  {author} {\bibinfo {author} {\bibfnamefont {G.~K.}\ \bibnamefont
  {Madsen}}\ and\ \bibinfo {author} {\bibfnamefont {D.~J.}\ \bibnamefont
  {Singh}},\ }\href {\doibase 10.1016/j.cpc.2006.03.007} {\bibfield  {journal}
  {\bibinfo  {journal} {Computer Physics Communications}\ }\textbf {\bibinfo
  {volume} {175}},\ \bibinfo {pages} {67} (\bibinfo {year} {2006})}\BibitemShut
  {NoStop}%
\bibitem [{\citenamefont {Madsen}\ \emph {et~al.}(2018)\citenamefont {Madsen},
  \citenamefont {Carrete},\ and\ \citenamefont {Verstraete}}]{BoltzTraP2}%
  \BibitemOpen
  \bibfield  {author} {\bibinfo {author} {\bibfnamefont {G.~K.~H.}\
  \bibnamefont {Madsen}}, \bibinfo {author} {\bibfnamefont {J.}~\bibnamefont
  {Carrete}}, \ and\ \bibinfo {author} {\bibfnamefont {M.~J.}\ \bibnamefont
  {Verstraete}},\ }\href {\doibase 10.1016/j.cpc.2018.05.010} {\bibfield
  {journal} {\bibinfo  {journal} {Comput. Phys. Commun.}\ }\textbf {\bibinfo
  {volume} {231}},\ \bibinfo {pages} {140 } (\bibinfo {year}
  {2018})}\BibitemShut {NoStop}%
\bibitem [{\citenamefont {Ederer}\ and\ \citenamefont
  {Spaldin}(2005{\natexlab{c}})}]{Ederer2005}%
  \BibitemOpen
  \bibfield  {author} {\bibinfo {author} {\bibfnamefont {C.}~\bibnamefont
  {Ederer}}\ and\ \bibinfo {author} {\bibfnamefont {N.}~\bibnamefont
  {Spaldin}},\ }\href {\doibase 10.1103/PhysRevB.71.060401} {\bibfield
  {journal} {\bibinfo  {journal} {Physical Review B}\ }\textbf {\bibinfo
  {volume} {71}},\ \bibinfo {pages} {060401} (\bibinfo {year}
  {2005}{\natexlab{c}})}\BibitemShut {NoStop}%
\end{thebibliography}
%

\end{document}